\documentclass[remotesensing,article,submit,pdftex,moreauthors]{Definitions/mdpi} 
\preto{\abstractkeywords}{\nolinenumbers}
\firstpage{1} 
\pubvolume{1}
\issuenum{1}
\articlenumber{0}
\pubyear{2022}
\copyrightyear{2022}
\datereceived{} 
\dateaccepted{} 
\datepublished{} 
\hreflink{https://doi.org/} % If needed use \linebreak
\makeatletter
\let\c@lofdepth\relax
\let\c@lotdepth\relax
\makeatother
\usepackage{subfigure}
\usepackage{bm}
\usepackage{xcolor}
\usepackage{textcomp}

\Title{A Systematic Approach of Inertial Sensor Calibrations for Gravity Recovery Satellites and Its Application to Taiji-1 Mission}

\TitleCitation{A Systematic Approach for Inertial Sensor Calibrations of Gravity Recovery Satellites and Its Application to Taiji-1 Mission}

\Author{Haoyue Zhang $^{1,2}$, Peng Xu $^{2,1,3,4}$*\orcidA{}, Zongqi Ye$^{1,2}$, Dong Ye$^{5}$\orcidB{}, Li-E Qiang$^{6}$\orcidC{}, Ziren Luo$^{2,3,4}$\orcidD{}, Keqi Qi$^{2}$, Shaoxin Wang$^2$, Zhiming Cai$^{7}$, Zuolei Wang$^8$, Jungang Lei$^{8}$, Yueliang Wu$^{3,4,9,10}$}

\AuthorNames{Firstname Lastname, Firstname Lastname and Firstname Lastname}

\AuthorCitation{H. Y. Zhang; P. Xu; Z. Q. Ye and et. al.}
\address{$^{1}$ \quad Lanzhou Center of Theoretical Physics, Lanzhou University, Lanzhou 730000, China. \\
$^{2}$ \quad Center for Gravitational Wave Experiment, Institute of Mechanics, Chinese Academy of Sciences, Beijing 100190, China.\\
$^{3}$ \quad Taiji Laboratory for Gravitational Wave Universe (Beijing/Hangzhou), University of Chinese Academy of Sciences (UCAS), Beijing 100049, China.\\
$^{4}$ \quad School of Fundamental Physics and Mathematical Sciences, Hangzhou Institute for Advanced Study, UCAS, Hangzhou 310024, China.\\
$^{5}$ \quad Harbin Institute of Technology, Harbin 150001, China.\\
$^{6}$ \quad National Space Science Center, Chinese Academy of Sciences, Beijing 100190, China.\\
$^{7}$ \quad Innovation Academy for Microsatellites, Chinese Academy of Sciences, Shanghai, China.\\
$^{8}$\quad  Lanzhou Institute of Physics, China Academy of Space Technology, Gansu 730000, China.\\
$^{9}$\quad  CAS Key Laboratory of Theoretical Physics, Institute of Theoretical Physics, Chinese Academy of Sciences, Beijing 100190, China.\\
$^{10}$ \quad International Centre for Theoretical Physics Asia-Pacific, UCAS, Beijing 100190, China.}

\corres{Correspondence: xupeng@imech.ac.cn}

\abstract{ High-precision inertial sensors or accelerometers can provide us references of free-falling motions in gravitational field in space.
They serve as the key payloads for gravity recovery missions such as the CHAMP, the GRACE-type missions, and the planned Next Generation Gravity Missions. 
In this work, a systematic method of electrostatic inertial sensor calibrations for gravity recovery satellites is suggested, which is applied to and verified with the Taiji-1 mission. 
{ With this method, the complete operating parameters including the scale factors, the center of mass offset vector and the intrinsic biased acceleration can be precisely calibrated with only two sets of short-term in-orbit experiments. This could reduce the data gaps that caused by necessary in-orbit calibrations during the lifetime of related missions.}  
Taiji-1 is the first technology demonstration satellite of the ``Taiji Program in Space'', which, in its final extended phase in 2022, could be viewed as operating in the mode of a high-low satellite-to-satellite tracking gravity mission.
Based on the calibration principles, swing maneuvers with time span about 200 s and rolling maneuvers for 19 days were conducted by Taiji-1 in 2022. 
{  Given the data of the actuation voltages of the inertial sensor, satellite attitude variations, precision orbit determinations, the inertial sensor's operating parameters are precisely re-calibrated with Kalman filters  and are updated to the Taiji-1 science team. The relative errors of the calibrations turn out to be $<1\%$ for the linear scale factors, $< 3\%$ for center of mass and $<0.1\%$ for biased accelerations.}
Data from one of the sensitive axis is re-processed with the updated operating parameters,  and the performance is found to be slightly improved compared with former results. 
This approach could be of high reference value for the accelerometer or inertial sensor calibrations of the GFO, the Chinese GRACE-type mission, and the Next Generation Gravity Missions. 
This could also shed some light on the in-orbit calibrations of the ultra-precision inertial sensors for future GW space antennas because of the technological inheritance between these two generations of inertial sensors. }


\begin{document}

\section{Introduction \label{Intro}}

Space-borne high-precision inertial sensors (IS) or accelerometers (ACC) can provide us references of  the inertial or free-falling motions in gravitational field in space, 
and therefore play the key role in satellite missions related to (Newtonian or relativistic) gravitational field measurements, especially for gravity recovery missions and gravitational wave detections in space. 
Considering the different implementation technologies, IS that based on the electrostatic suspension and servo control technologies is still of the most precise and reliable inertial payload in present days.
Electrostatic IS when working in the ACC mode can measure precisely the non-gravitational forces that exert on the satellites, and had already served the series of gravity recovery missions since the beginning of this century, including the CHAMP \cite{reigber2002champ}, GRACE/GFO \cite{davis1999grace, tapley2004grace, flechtner2014gfz,dahle2019gravity}, GOCE \cite{drinkwater2003goce} and also Taiji-1~\cite{yue2021china,wu2022global} missions.
Aided by the technology of drag-free controls, electrostatic inertial sensors could reach the unprecedented ultra-precision level  ($\sim 10^{-15}m/s^2/Hz^{1/2}@3mHz$) of inertial motions in space, which was successfully demonstrated by the LISA PathFinder mission \cite{mcnamara2019lisa}.
Such ultra-precision IS will be the key payloads
% that can serve as the end mirrors of the inter-satellite laser interferometers, 
of future Gravitational Wave (GW) antennas in space, including LISA \cite{amaro2017laser}, Taiji \cite{hu2017taiji} as well as TianQin~\cite{luo2016tianqin}.

For the high-low (such as  the CHAMP and Taiji-1 missions) or low-low (like the GRACE and GFO missions) satellite-to-satellite tracking gravity missions, the details of Earth geopotentials are encoded in the orbital motions or relative motions of satellites. 
To map out precisely the global gravity field, the high-precision and in-orbit measurements of the non-gravitational forces that perturb the satellite orbits are required.
For the present-day gravity missions including CHAMP, GRACE/GFO, this { was} achieved by on-board electrostatic IS systems that working in the ACC mode.
The designs and the working principles for these IS systems are basically the same. 
The IS contains a Test Mass (TM) suspended inside an electrode cage as the reference of inertial motions, and a front end electronics (FEE) unit to read out and adjust the relative motions between the TM and the cage.
In the ACC mode, the compensation or actuation voltages that { push and maintain} the TM back to its nominal position will give rise to the precision measurements of the non-gravitational perturbations on the satellite (see detailed explanations in subsection \ref{sec2.2}).
Such ACC data is then included in the modelling of the satellite orbits and {  fitting} of the Earth geopotentials. 
For GOCE and concepts of the Next Generation Gravity Mission (NGGM), the electrostatic IS could also work in the { drag-free} mode, that the non-gravitational perturbations on the satellite could be measured by the FEE of the IS but compensated by pushing, with $\mu$-N thrusters, the satellite to follow the inertial motions of the TM.
 For both the ACC and drag-free modes, the controlled dynamics of the TM relative to the cage is determined by the combined action of the non-gravitational forces on the spacecraft, weak disturbances on the TM, and also the compensation forces from the control loop.
 Therefore, to accurately interpret and make use of the IS data in gravity inversions, the parameters involved in defining the characteristics of the device need to be carefully measured and calibrated.
 These include, generally, the scale factors of each axis that transform the control voltages imposed on the TM into the non-gravitational forces exerted by the satellites; the bias voltages or accelerations in the readouts that come from the environmental DC forces and imbalances of the FEE; and also the offset vector from the {  center of mass} of the satellite which gives out to the confusing inertial accelerations in the readouts.

 In 2022, the Taiji-1 was in its final extended phase and operated in the high-low satellite-to-satellite tracking mode,
 and during this year radical experiments for Taiji-1 were performed including especially the monthly global gravity field recovery experiment \cite{wu2022global} and the re-calibrations of the key measurement system (laser interferometer and IS) with satellite maneuvers after its two years operation. 
%Taiji-1 is the first technology demonstration satellite of the ``Taiji Program in Space'', which was approved by the Chinese Academy of Science (CAS) in { August} 2018.
The Taiji-1 IS system has similar designs as those onboard  CHAMP~\cite{rehm2000champ}, GRACE/GFO~\cite{ries2001design}, GOCE~\cite{floberghagen2013gravity}, etc, and 
can work in both the ACC mode and the drag-free control mode.
%  after the phase A study and the ground-based tests of related technologies~\cite{liu2018principle, liu2018development, li2018laser, li2019demonstration,li2020laser, liu2021numerical}. 
 The ``Taiji Program in Space'', initiated to expert demonstration in 2008 and officially released by the CAS in 2016, is for China's space-borne GW observatory, namely the Taiji mission \cite{hu2017taiji,luo2021taiji,luo2020brief,ruan2020taiji}. 
% Taiji will launch three spacecrafts into heliocentric orbits that forming a nearly equilateral triangular constellation with arm length $\sim 3$ million kilometers, and could enclose the GW sources as coalescing supermassive black hole binaries, extreme mass ratio inspires, stochastic gravitational wave backgrounds, etc..  
% According to Taiji's road map \cite{luo2021taiji,luo2020brief}, it is expected that the science operations of LISA and Taiji may overlap in the 2030s and a network of space antennas would be formed \cite{wang2021alternative,omiya2020searching,ruan2020lisa,ruan2021lisa}.    
% The assembly integrations and tests of Taiji-1 started from APR 2019, and in AUG 2019 Taiji-1 was launched.
The successful operation of Taiji-1 in 2019 and 2020 had demonstrated and confirmed the designed performances of the scientific payloads and the satellite platform, and verified the most important individual technologies of China's space GW antenna and possible gravity missions, including high precision laser interferometers, drag-free control system, $\mu$-N thrusters, the ultra-stable and clean platform, and especially the electrostatic inertial sensor~\cite{yue2021china,wang2021development}.

Based on the classical works~\cite{wang2003study,van2009champ,armano2016sub,baghi2022detection,huang2022estimation,armano2018calibrating,armano2015bayesian,bezdvek2010calibration,klinger2016role,behzadpour2021grace} and especially the valuable experiences of the calibration experiments of the GRACE and GFO missions~\cite{wang2003study,van2009champ,
bezdvek2010calibration,klinger2016role,behzadpour2021grace}, a new and systematic method for the complete operating parameters calibration of electrostatic IS system for gravity missions, including the scale factors, the acceleration biases, the offset of { the Center of Mass (COM)}, was suggested by us and approved by the Taiji science team in the end of  2021. 
During 2022, a set of satellite maneuvers for Taiji-1 IS calibration were conducted, including the high frequency swings (period $\sim 30\ s$) and rollings (period $\sim 724 \ s$) of the satellite along certain axes. 
With the observational data from the IS, star trackers and precision orbit determinations during the calibration phase, the IS operating parameters are determined with {  high accuracy} to study their possible variations and drifts during the two years operations.  
Compared with the original values that determined by ground-based and in-orbit experiments \cite{wang2021development, cai2021satellite}, variations of these parameters can be identified, which might be caused by the mechanical disturbances during the launch, changes of the {  center of mass} of the spacecraft due to consumption of cold gas, and the aging of the electronics unit in the past three years.
Based on such re-calibrated parameters, we revisit the pre-processing of the IS data, and the resolution of one of the sensitive axes of Taiji-1 IS (the z-axis) is found to be slightly improved in the interested frequency band. 
Such a systematic approach could be applied to the ACC or IS calibrations of gravity recovery missions like the GFO, the Chinese GRACE-type mission, and the future planned Next Generation Gravity Missions.  
Moreover, this approach could also shed some light on the in-orbit calibrations of the ultra-precision IS for future GW space antennas, since the principle and technology have inheritance between these two generations of the electrostatic IS payloads.

This work expands as follows: In Sec. \ref{IS} we briefly introduce the Taiji-1 satellite and the IS payload. 
The requirement of the IS calibrations and the systematic method we adopted are described in Sec. \ref{method}. In Sec. \ref{RES}, the observational data and the processing procedure are introduced, and the re-calibrated operating parameters are compared with the original used ones; furthermore, we re-process the representative data of the Taiji-1 IS and re-evaluate its performance. The conclusion of this work could be found in Sec. \ref{Conclu}.

%%%%%%%%%%%%%%%%%%%%%%%%%%%%%%%%%%%%%%%%%%%%%%%%%%%%%%%%%
 
\section{Inertial Sensor of Taiji-1\label{IS}}
\subsection{Taiji-1 satellite}
{According to the three step road map of the Taiji program \cite{luo2021taiji,luo2020brief}, %Taiji-1 is the first technology demonstration satellite. 
 the Taiji-1 satellite weights about 180 kg, and the key measurement system contains the drag-free control system and the optical metrology system.}  
Taiji-1 was launched to a circular dawn/dusk Sun-synchronous orbit, with the altitude about $600\ km$ and inclination angle $97.67^{o}$.
The orbit has a stable Sun-facing angle, which can provide a constant power supply for the battery and also the stable temperature gradient for the platform. 
The orbit coordinates system is defined as follows, the $+X$-direction is along the flight direction, $+Z$ the radial direction and $+Y$ is defined by the right-hand rule.

\begin{figure}[htbp]
\centering
\includegraphics[scale=0.45]{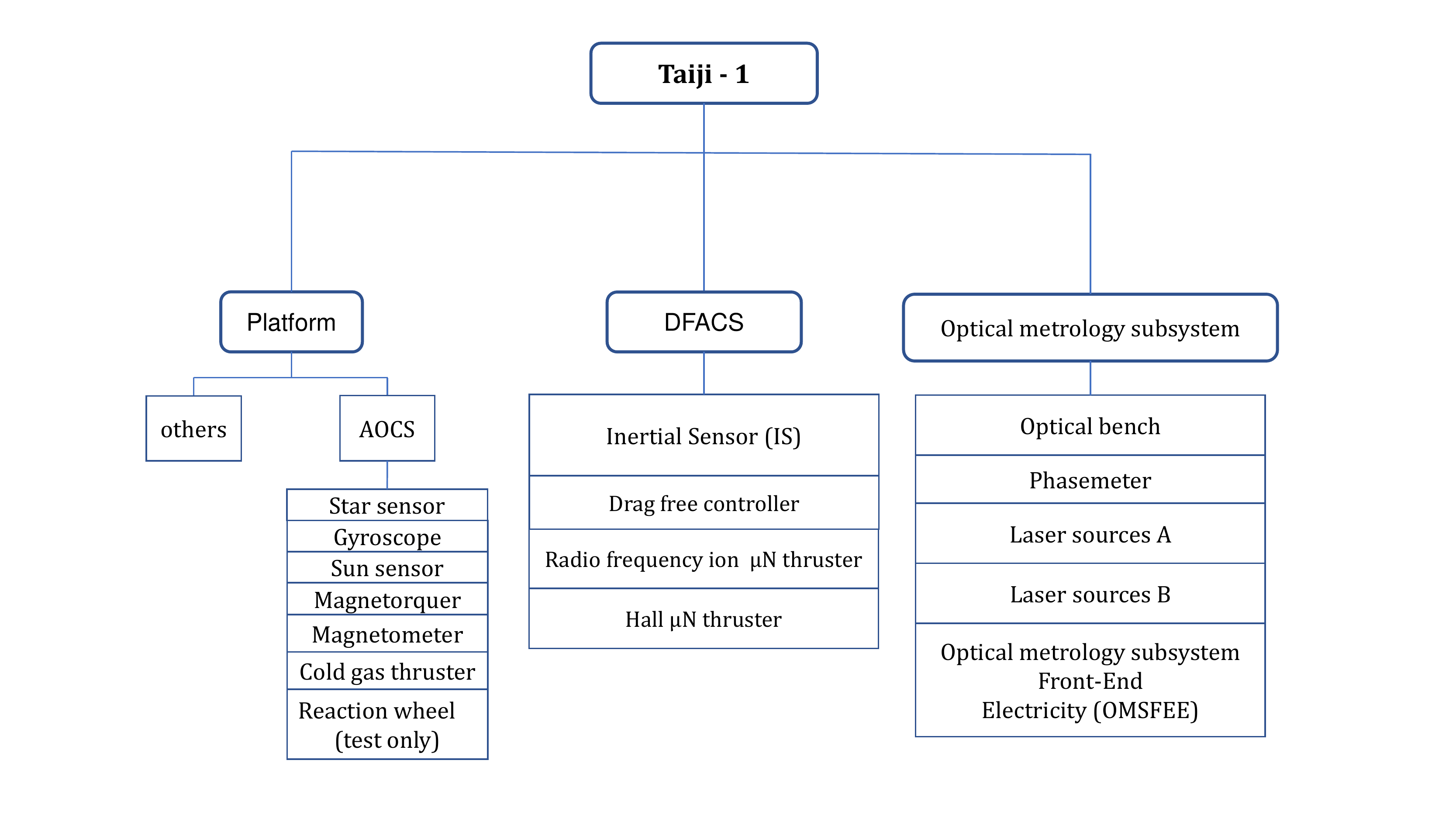}
\caption{Block diagram for the architecture of Taiji-1.}
\label{fig:Taiji-1_archi}
\end{figure}

\begin{figure}
    \centering
    \includegraphics[scale=0.46]{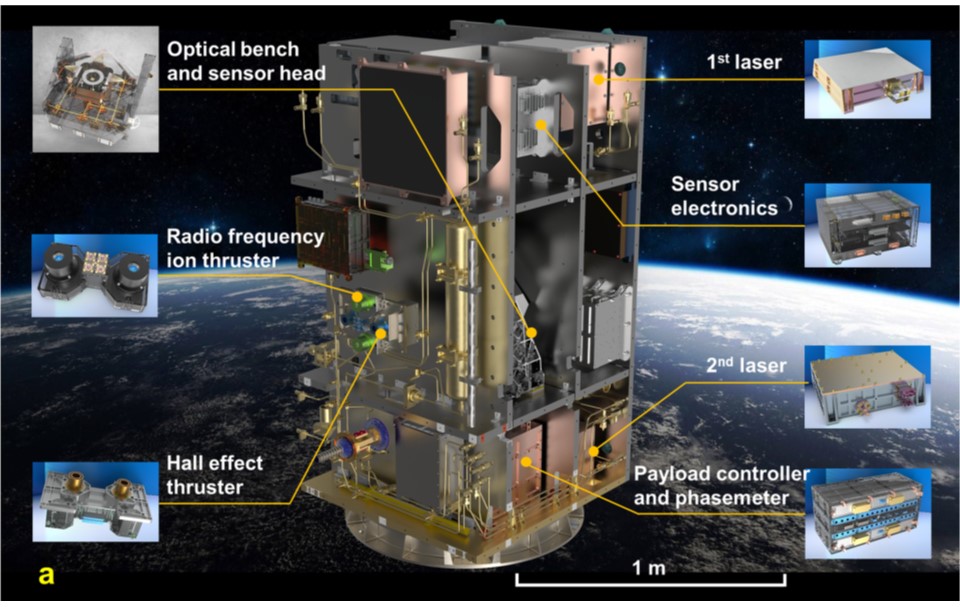}
    \caption{The layout of the payloads of Taiji-1.}
    \label{fig:payloads}
\end{figure}

The IS was installed at the {  center of mass} of the spacecraft (SC), having the nominal offset $\leq 150\ \mu m$. 
The IS, the drag-free control unit, and the two types of micro-thrusters, that the Hall and radio frequency iron $\mu$N thrusters, together constitute the drag-free control system. 
The optical metrology system contains an optical bench, a high-precision phasemeter and two Na-YAG laser sources. 
The TM interferometer can provide the independent readout of the position of the TM in the $x$-axis of the IS and the optical bench interferometer serves as a reference.
Both the interferometers reached the resolutions $\leq 100pm/Hz^{1/2}$ \cite{yue2021china}.
The ultra-stable and clean satellite platform has a highly stable thermal control system, which provides the $\sim \pm 1200\ mK$ thermal stability of the satellite environment, $\sim \pm 350 \ mK$ in the middle cabin and about $\pm 2.6 \ mK$ for the key measurement system \cite{yue2021china,cai2021satellite}.
The reaction wheel installed along the Y-direction of the satellite is the only movable unit, which is for reliability considerations. 
The attitude and orbit control system contains the star trackers, gyroscopes, sun sensor, magnetometer, magnetorquer, cold gas thrusters and the controller. 
The architecture of Taiji-1 is shown in Fig. (\ref{fig:Taiji-1_archi}) and Fig. (\ref{fig:payloads}).

\subsection{Inertial sensor} \label{sec2.2}

The electrostatic IS system of Taiji-1 contains mainly the mechanical assembly, the FEE
unit, and auxiliary subsystems like vacuum chambers, etc. 
The mechanical assembly {  consists of} a 72 g and $4\ cm \times 4\ cm \times 1\ cm$ parallelepipedic TM of titanium alloy and an electrode cage made of ultra low expansion silica that encloses the TM.

\begin{figure}[h]
\centering
\includegraphics[scale=0.33]{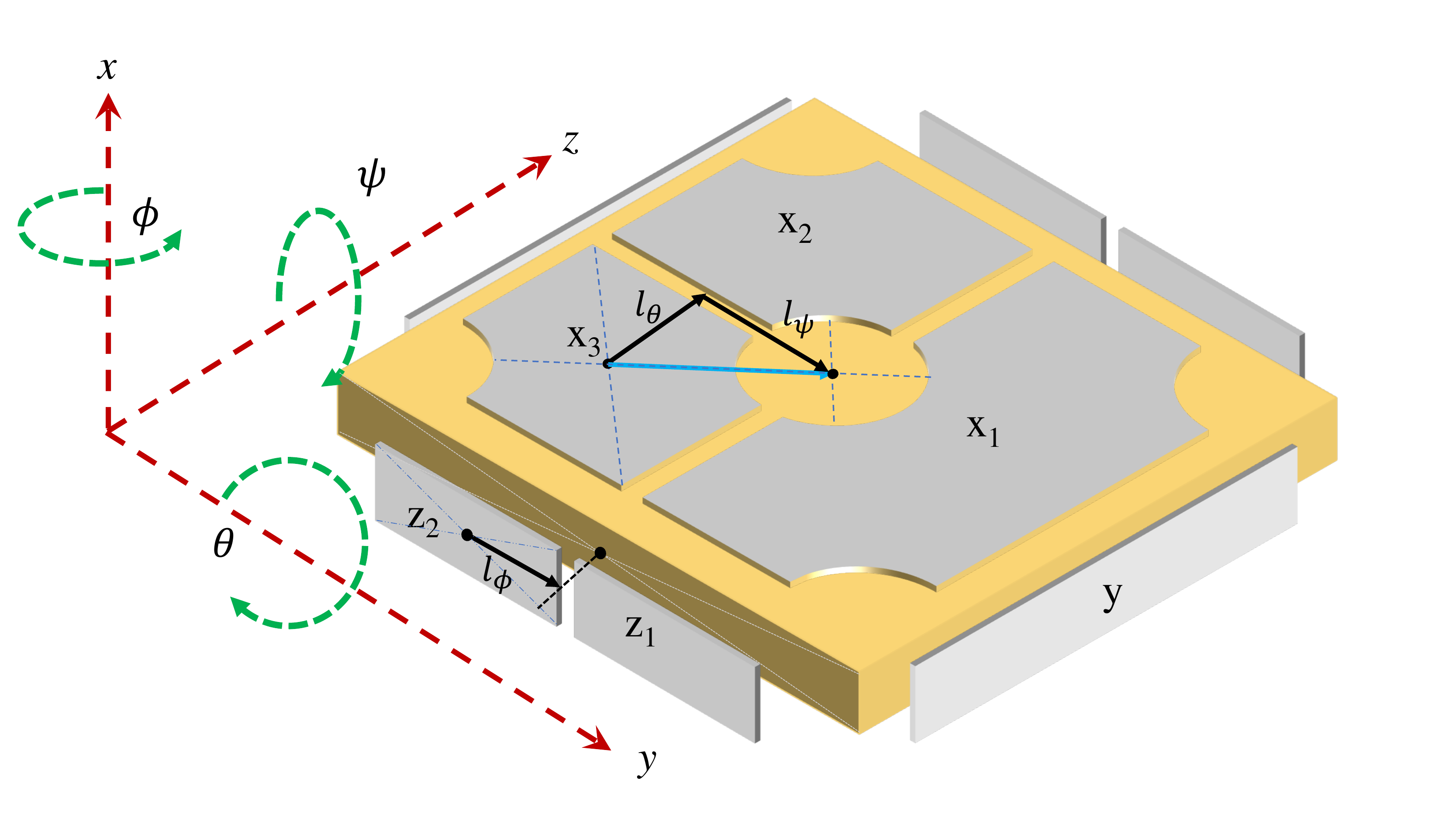}
\caption{layout of the core mechanical assembly of the IS.}
\label{fig:IS}
\end{figure}

Both the TM and the cage are gold coated, and inside the cage there are six pairs of electrodes facing the TM side faces, see Fig. (\ref{fig:IS}) for illustration and the definitions of the measurement axes in the IS frame.
The TM, serves as the inertial reference, is suspended electrostatically inside the cage. 
When operating, the position variations of the TM relative to the cage { causes changes} of the capacitance between the TM side faces and electrodes, which induces signals through Wheatstone bridges that can be picked up by the FEE to give rise to the measurements of the TM position and attitude. 
Based on such data and the PID algorithm, the TM is servo-controlled to its nominal position by applying low frequency actuation voltages through the same electrode in the accelerometer mode or by pushing the spacecraft through $\mu N$-thrusters in the drag-free control mode. 
In the normal science operations of Taiji-1, the +z direction of the IS points to the flight direction, the +x the radial direction, and +y is defined by the right-hand rule.
{ Along with Fig. \ref{fig:IS}, the mechanical and geometrical parameters of the mechanical assembly used in the following sections are listed in Tab.\ref{tab:TM}, and their detailed definitions can be found in subsection \ref{subsec:SF&COM principle}. }
\begin{table}[htbp]
    \begin{center}
    \caption{{ Parameters related to the structures of the mechanical assembly.}}
    \label{tab:TM}
    \begin{tabular}{c|c|c|c|c}
    \hline
     Parameter &   x-axis($\phi$) &   y-axis($\theta$) &   z-axis($\psi$)& units\\
    \hline
    $l$   &    $8.00\times 10^{-3}\ $&$10.45\times 10^{-3}\ $ &$10.45\times 10^{-3}\ $ & $m$\\
    \hline
    $J$  &     $1.9440\times 10^{-5}\ $&$1.0328\times 10^{-5}\ $ & $1.0328\times 10^{-5}\ $ & $kg\cdot m^2$\\
    \hline
    $D$   &    $6.20\times 10^{-5}\ $&$7.70\times 10^{-5}\ $ &$7.50\times 10^{-5}\ $ & $m$\\
    \hline
    $S$   &    $2.395\times 10^{-4}\ $&$2.054\times 10^{-4}\ $ &$2.056\times 10^{-4}\ $ & $m^2$\\
    \hline
    \end{tabular}
    \end{center}
    \end{table}

The IS science and housekeeping data is archived and processed at the Taiji-1 data processing center of CAS in Beijing, and the data management and the detailed processing flows can be found in \cite{jin2021pipeline}. 
The level 1 science data product contains the actuation voltages on the six electrodes. 
With the calibrated scale factors and biases, the actuation voltages are transformed into linear accelerations and angular accelerations of the TM relative to the cage, and with the COM offset corrected these are then written into the level 2 science data product. 
The position sensor data together with the IS state monitoring data including temperatures of the core assembly and the FEE unit, biased reference voltages are all packaged as IS housekeeping data product. 

The nominal precision level (or the acceleration noise level) of Taiji-1's IS is $3\times10^{-9}\ m/s^{2}/Hz^{1/2}@10\ mHz$,  see Tab. \ref{tab:IS}  for the key design requirements. 
The in-orbit performance was evaluated with the measurement of the y-axis, which is one of the sensitive axes pointing to the orbital normal direction. 
The amplitude spectrum density (ASD) of the acceleration measured by the y-axis was $\leq 2\times10^{-9}\ m/s^{2}/Hz^{1/2}$ \cite{yue2021china,wang2021development}, that fulfilled the design requirements. 
The IS couples to the space environment and satellite platform in a rather complicated way. 
To be more specific, the dynamics of the TM or the residual acceleration ($a_{R}^{i}(t), i=x,\ y,\ z$) of the TM relative to the platform can be written down as 
\begin{align}
a_{R}^{i}(t) &=  a_{TM}^{i}(t)-a_{SC}^{i}(t)\nonumber \\
 &=  a_{grav,TM}^{i}(t)+a_{para,TM}^{i}(t)-a_{grav,SC}^{i}(t)-a_{para,SC}^{i}(t)+a_{in,TM}^{i}(t)+a_{c}^{i}(t-\tau_c^{i})\nonumber \\
 &=  a_{c}^{i}(t-\tau_{c}^{i})-a_{para,SC}^{i}(t)+G^{ij}(t)d_j+a_{para,TM}^{i}(t).\label{eq:aACC}
\end{align}
In the first line, $a_{TM}^{i}(t)$ and $a_{SC}^{i}(t)$ denote the acceleration of the TM and spacecraft with respect to the local inertial frame. 
In the second equal, the accelerations could be expanded as the followings.
$a_{grav,SC}^{i}(t)$ and $a_{grav,TM}^{i}(t)$ are the gravitational accelerations of the SC and TM respectively. 
$a_{para,SC}^{i}(t)$ and $a_{para,TM}^{i}(t)$ are of parasitic accelerations, where $a_{para,SC}^{i}(t)$ mainly comes from non-gravitational forces from solar radiation, air drag, and Earth albedo acting on the satellite and also the mechanical disturbances from the satellite platform, and $a_{para,TM}^{i}(t)$ are from noise forces acting directly on the TM including actuation noises, spring-like couplings, radiometric effect, magnetic couplings, thermo-noises from gold wire attached to the TM and etc. 
$a_{in,TM}^{i}(t)$ is the inertial acceleration that comes from the relative attitude variation between the TM and the satellite.
$a_{c}^{i}(t-\tau_{c}^{i})$ is the compensation acceleration to keep the TM to its nominal position, and $\tau_{c}$ is the delay time of the control loop. In the ACC mode $a_c^i(t)$ is the electrostatic actuation force acting on the TM, while in the drag-free control mode $a_c^i(t)$ is the compensation force by the $\mu N$-thrusters acting on the satellite. 
At last, in the third line, the gravitational gradients and the inertia acceleration can be summarized into the term that is proportional to the COM offset $d_i$
\begin{equation}
    G^{ij}(t)d_j=T^{ij}(t)d_{j}+\omega^{ik}(t)\omega_{k}^{\ j}(t)d_j+\dot{\omega}^{ij}(t)d_{j} \label{eqn:com1},
\end{equation}
where $T^{ij}$ denotes the components of the gravitational tidal matrix and $\omega^{ij}$ the angular velocity matrices of the satellite relative to the local inertial frame.

\begin{table}[htbp]
\begin{center}
\caption{Key design requirements of Taiji-1 IS system.}
\label{tab:IS}
\begin{tabular}{c|c}
\hline
Parameter &   Nominal Value\\
\hline
Dynamic range    &    $\sim 3\times 10^{-5}\ m/s^2$\\
\hline
Bandwidth      &      $10\ mHz \sim  1\ Hz$\\
\hline
Position noise   &   $\leq 1\ nm/Hz^{1/2}$ \\
\hline
Actuation noise  &   $\leq 2\times 10^{-5}\ V/Hz^{1/2}$\\
\hline
Readout noise & $\leq 2\times 10^{-5}\ V/Hz^{1/2}$\\
\hline
Acceleration noise & $\leq 3\times10^{-9}\ m/s^{2}/Hz^{1/2}$\\
\hline
\end{tabular}
\end{center}
\end{table}

In this work, we consider the ACC mode, the compensation force $a_c^i(t)$ is read out in terms of the actuation voltages and is of the science data of the IS system.
\begin{eqnarray}
a_c^i(t) & = & a_{para,SC}^i(t)-a_{para,TM}^i(t)-G^{ij}(t)d_j+a_R^i(t),\label{eq:ac}
\end{eqnarray}
where the actuation acceleration reads
\begin{eqnarray}
a_c^i(t) & = & b^i+k^{i\alpha}_1 V_{\alpha}(t)+k^{i\alpha\beta}_2 V_{\alpha}(t)V_{\beta}(t).\label{eq:acV}
\end{eqnarray}
Here $V_{\alpha},\ (\alpha=x1,\ x2,\ x3,\ y,\ z1,\ z2)$ are actuation voltages on each electrode, $b^i$ denotes the acceleration bias, $k^{i\alpha}$ the linear scale factors that transform the voltages to accelerations, and $k^{i\alpha\beta}$ the quadratic factors. 
In the normal science mode of the IS, the TM is controlled to tightly follow the motions of the electrode cage or the satellite platform, that their relative motions are $\leq 10^2 pm/Hz^{1/2}$ in the sensitive band. 
This means that the residual acceleration term $a_R^i(t)$ in Eq. (\ref{eq:ac}) can be ignored. 
The other term that can be ignored in practical use is the quadratic term $k^{i\alpha\beta}_2 V_{\alpha}(t)V_{\beta}(t)$ in Eq. (\ref{eq:acV}).
Therefore, with the COM offset term $G^{ij}(t)d_j$ been corrected, and the scale factors, bias determined, the voltages data or the actuation accelerations data can give rise to the measurements of the non-gravitational forces exerted by the satellite,
\begin{equation}
    a_{para,SC}^i(t)=k_1^{i \alpha}V_{\alpha}(t)+G^{ij}(t)d_j+a_{para,TM}^i(t)+b^i.\label{aSC}
\end{equation}
The parasitic acceleration noises $a_{para,TM}^i(t)$ acting on the TM determines the noise floor of the IS system.

%%%%%%%%%%%%%%%%%%%%%%%%%%%%%%%%%%%%%%%%%%%%%

\section{Principle of IS calibration \label{method}}

As discussed in the previous section, to correctly interpret and make use of the IS data, one needs to carefully determine the operating parameters of the IS device. 
Even though some of the relevant parameters were calibrated with ground based experiments before launch or in-orbit experiments in the commissioning phase,  large disturbances during the launch, consumption of {  consumable gas}, agings of the electric units and etc. may still { cause changes} of the characteristic of the IS device.
Therefore, for Taiji-1's IS system and missions that { carrying similar} electrostatic IS payloads, it is necessary to calibrate the basic set of operating parameters, including the scale factors $k^{i\alpha}$, linear bias $b^i$, and the COM offset $d_i$, with the in-orbit data and regularly within the mission lifetime.

{ In the following}, we discuss the calibration principles of this set of parameters and the related satellite maneuver strategies, that are adopted for Taiji-1's calibration. 
The key considerations here are to try to complete the IS calibrations with less satellite maneuvers and shorter calibration time durations, and also try to reduce the possible risks as much as possible.

\subsection{principle of scale factors and COM offset calibrations \label{subsec:SF&COM principle}}

For electrostatic IS systems with parallelepipedic TMs such as the cases for Taiji-1, GRACE/GRACE-FO and etc, the scale factors appeared in Eq. (\ref{eq:acV}) can be divided into two sets, that the linear scale factors $[k^x,\quad k^y,\quad  k^z]$ and angular scale factors $[\beta^x,\quad  \beta^y,\quad  \beta^z]$, which transform the actuation voltages imposed on the electrodes into the corresponding compensation linear accelerations $a_c^i$ and angular accelerations $\dot{\omega}^i_c$ respectively. 
For Taiji-1, given the geometrical and mechanical parameters of the TM and the electrodes, the nominal values of the two sets of scale factors can be derived,
\begin{align}
    k^x&=\frac{2\epsilon_0S_xV_P}{MD_x^2}, \quad
    k^y=\frac{2\epsilon_0S_yV_P}{MD_y^2},  \quad
    k^z=\frac{2\epsilon_0S_zV_P}{MD_z^2}, \label{eq:k} \\ 
    \beta^x&=\frac{\epsilon_0S_zl_{\phi}V_P}{J_{\phi}D_z^2},\quad
    \beta^y=\frac{\epsilon_0S_xl_{\theta}V_P}{J_{\theta}D_x^2},\quad 
    \beta^z=\frac{\epsilon_0S_xl_{\psi}V_P}{J_{\psi}D_x^2}. \label{eq:beta}
\end{align}
Here $M$ stands for the mass of the TM, $J_{\phi}$, $J_{\theta}$, $J_{\psi}$ denote the mass moment of the TM along the $x$, $y$ and $z$ axes.
$S_i$ is the total area of electrode surface of the \emph{ith} axis, $D_i$ the nominal distance between the TM surface and the electrode, and $l_{i}$ denotes the force arm of the electrodes pair that control the $ith$ rotation degree of freedom, see again Fig. \ref{fig:IS}. 
$\epsilon_0$ stands for the vacuum permittivity, and $V_P$ the preload biased voltage. { The values of these parameters for Taiji-1 are shown in Tab.\ref{tab:TM}}, and the transformation relations between the actuation voltages and compensation accelerations are shown in Tab. \ref{table1}.

\begin{table}[htbp]
\begin{center}
\caption{The actuation accelerations, angular accelerations and the corresponding  scale factors.\label{table1}}
\begin{tabular}{c|c}
\hline
Compensation acceleration readouts &  Scale factors and actuation voltages \\
\hline
$a_c^x$&$k^x(2V_{x1}+V_{x2}+V_{x3})$\\
$a_c^y$&$k^yV_{y}$\\
$a_c^z$&$k^z(V_{z1}+V_{z2})$\\
$\dot{\omega}_c^x$&$\beta^x(V_{z1}-V_{z2})$\\
$\dot{\omega}_c^y$&$\beta^y(V_{x2}-V_{x3})$\\
$\dot{\omega}_c^z$&$\beta^z(2V_{x1}-V_{x2}-V_{x3})$\\
\hline
\end{tabular}
\end{center}
\end{table}

According to the designs of the Taiji-1's IS system, we have the following useful relations in calibrating the linear scale factors and the COM offset,
\begin{align}
    \beta^x&= \frac{Ml_{\phi}}{2J_{\phi}} k^z\label{e7}, \\ 
    \beta^y&= \frac{Ml_{\theta}}{2J_{\theta}} k^x\label{e8}, \\
    \beta^z&= \frac{Ml_{\psi}}{2J_{\psi}} k^x\label{e9}.
\end{align}
These relations remained unchanged during the mission lifetime since they involve only the geometrical and mechanical properties of the TM and the electrode cage. 
The high machining accuracy ($\delta l/l \sim 10^{-4}$) of the TM and cage structures ensures that the relations between the scale factors are sufficiently accurate.
In this case, that $l_{\theta} = l_{\psi}$ and $J_{\theta} = J_{\psi}$, we have { $\beta_y = \beta_z \label{byeqbz}$.}
Another important property is that, during the normal science operation of the IS in its ACC mode, the TM is controlled to tightly follow the motions of the electrode cage or the spacecraft. 
For Taiji-1, the position fluctuations of the TM relative to each electrode surface is $\leq 10^2 pm/Hz^{1/2}$ in the sensitive band.
This means that the rotations of the TM and the spacecraft could be treated as precisely synchronized,  that one has
\begin{align}
    \vec{\omega}_{TM}= \vec{\omega}_{SC},\qquad
   \vec{\dot{\omega}}_{TM}=\vec{\dot{\omega}}_{SC}.\label{e11}
\end{align}
Here, $\vec{\omega}_{TM}$, $\vec{\dot{\omega}}_{TM}$ and   $\vec{\omega}_{SC}$, $\vec{\dot{\omega}}_{SC}$ denote the angular velocities and angular accelerations of the TM and the spacecraft respectively.

Therefore, despite the offset between the installation orientations of the IS system and the star trackers, the measured angular velocities and accelerations of the spacecraft and the TM are interchangeable.

\begin{figure}[h]
    \centering
    \includegraphics[scale=0.5]{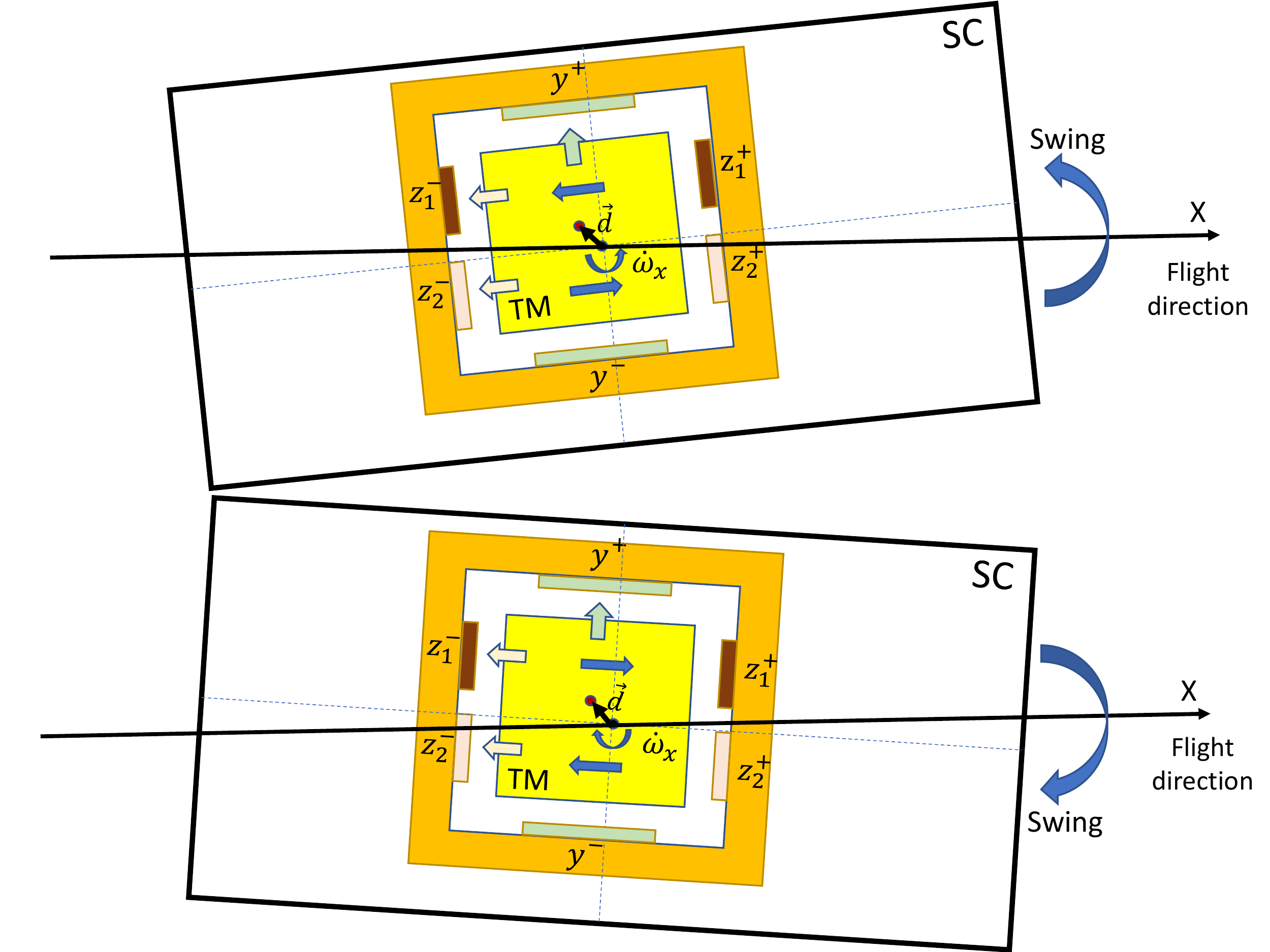}
    \caption{Satellite swing maneuver for the calibrations of the scale factors and COM offset.}
    \label{fig:swing}
    \end{figure}

The rotations or attitude variations $\vec{\omega}_{SC}$ and $\vec{\dot{\omega}}_{SC}$ of the spacecraft could be independently measured by on-board star trackers.
This motivates us to make use of such attitude measurements to calibrate the scale factors, which is different from the former methods based on precision orbit determination (POD) data \cite{van2009champ, hong2017new}.   
One could swing the spacecraft periodically along certain axis with a rather large angular accelerations and with relative higher frequency compare to the signals band of air drags and Solar radiations, that could be clearly identified and precisely measured by the IS system.
With the inputs of the angular accelerations derived by the star track measurements and the actuation voltages readout by the front end electric unit of the IS system, one can fit the angular scale factors $\beta^i$ based on the equations in Tab. \ref{table1} with the least squares estimation or the Kalman filter algorithms. 
According to the relations between angular and linear scale factors in Eq. (\ref{e7}) - (\ref{e9}), the linear scale factors can be further determined.
For Taiji-1' IS, controls along the y-axis is independent of other degrees of freedom, its actuation voltage does not involve with any rotation controls of the TM.
Therefore, the linear scale factor $k^y$ for Taiji-1' IS system can not be calibrated with this method and is left for blank in this work.
Please see Fig. \ref{fig:swing} for the illustration of this calibration method.

For the COM offset calibrations, one notices that, according to Eq. (\ref{eqn:com1}) and  (\ref{eq:ac}), the periodic swing of the spacecraft will also couple to the COM offset and produce periodic linear accelerations along the axes that are perpendicular to the rotation axis due to inertial effects, see again Fig. \ref{fig:swing}.
According to Tab. \ref{table1}, one can then use the common mode readouts of the actuation voltages of each axis instead of the differential mode used in the scale factor calibrations, together with the spacecraft attitude data by the star trackers or the IS readouts itself to fit and calibrate the COM offset vector. 
Such method had been carefully studied and employed by the GRACE and GRACE-FO team \cite{wang2003study, davis1999grace, christophe2015new, flury2008precise}.
While, one notices that a possible interference may come from the gravity gradient signals, since the spacecraft attitude variations would also produce  periodic projections of the local gravity tidal force with the same frequency.
This, on the contrary, forces us to choose satellite maneuvers with small magnitude of attitude variations. 
In fact, for Taiji-1 and GRACE-type missions, the magnitudes of gravity gradients $\sim 10^{-6} /s^2$.
Therefore, according to Eq. (\ref{eqn:com1}), for COM offset $\lesssim 10^{-3}\ m$,  attitude variations $\delta \omega \sim 10^{-3}\ rad$  will give rise to interference signals $\lesssim 10^{-12}\ m/s^2$, which could be safely ignored.
{ However, to obtain larger calibration signals with small magnitude of attitude variations, one is then forced to swing the spacecraft with high frequencies that increases the magnitude of the last term $\dot{\omega^{ij}}d_j$ in Eq. (\ref{eqn:com1}).
If the above considerations are satisfied, the remained interference from the gravity gradients together with no-gravitational disturbances can be treated as a linear term due to the orbit evolutions and be fitted out and removed in data processing.}

For clarity,  based on  Eq. (\ref{eq:ac}) one can re-write the observation equations for the COM calibration as  
\begin{equation}
    \vec{a}_{c}(t)= \mathbf{A}(t)\vec{d}+\vec{a}_I t+\vec{b}_I \label{eq_com2},
\end{equation}
where
\begin{equation}
 \mathbf{A} =\left(
\begin{array}{ccc}
    -\omega_{y}^{2}-\omega_{z}^{2} & \omega_{x}\omega_{y}-\dot{\omega}_{z} & \omega_{x}\omega_{z}+\dot{\omega}_{y} \\
    \omega_{x}\omega_{y}+\dot{\omega}_{z} & -\omega_{x}^{2}-\omega_{z}^{2} & \omega_{y}\omega_{z}-\dot{\omega}_{x} \\
    \omega_{x}\omega_{z}+\dot{\omega}_{y} & \omega_{y}\omega_{z}+\dot{\omega}_{x} & -\omega_{z}^{2}-\omega_{x}^{2} 
\label{eqn_A}
\end{array}
\right).
\end{equation} 
Here,  $\vec{a}_I t+\vec{b}_I$ are the linear term from the non-gravitational accelerations acting on the spacecraft and the gravity tidal accelerations coupled to the TM.  
Given the swing maneuvers discussed above, the COM offset vector $\vec{d}$ could then be fitted out.

To summarize,  we suggest to  calibrate the IS scale factors and COM offset with one round swing maneuver of the Taiji-1 satellite.
To enhance the signal-to-noise ratio (SNR) and reduce the possible interferences, the swing maneuver should be of high frequency compared with the signal band of non-gravitational forces, and the swing amplitude should be small to  reduce the interference signals from gravitational tidal forces.  
Also, the time span of the maneuver should be short to make sure that the linearity of the tidal force model remains accurate enough.
At last but not least, the attitude maneuvers should not be driven by thrusters, since beside the disturbances caused by propulsion the misalignment of the thrusters could produce large interference signals in linear accelerations. 
With these considerations, the maneuvers conducted by Taiji-1 were swings of the satellite driven by the magnetic torquers along certain axis with period about $25 \sim 30\ s$ and total time spans $< 300\ s$.
To enlarge the angular acceleration, we operated the magnetic torquers at their full powers, and the wave-trains of the satellite angular velocity were triangular waves with  magnitudes about $1\times 10^{-4}\ rad/s$. 
The swing maneuvers were conducted on 18 MAY 2022, see Fig. \ref{fig:pre3} and \ref{fig:V_z} for illustrations, and the data processing and fittings are discussed in the next section.

\subsection{principle of IS bias calibration \label{subsec:bias principle}}

From the physical point of view, the intrinsic bias $b^i$ in the actuation acceleration measurements in Eq.  (\ref{aSC}) mainly comes from the asymmetry of the electrodes on the opposite sides of the same axis and the imperfection in FEE unit. 
The imbalance of mass distributions surrounding the IS system, couplings between the TM and residual magnetic field and etc., may also contribute to the intrinsic bias accelerations.
Therefore, the intrinsic bias along each axis is stable, and its changes could be ignored in short time measurements. 
On the contrary, the projections of the DC or very low frequency non-gravitational forces along each axis change not only with the orbit positions but also with the attitude of the satellite. 

Generally, the long term energy loss due to orbital decays based on the POD data and the work done by the drag forces evaluated by the IS data need to be balanced, which gives us a method to find out the intrinsic biases. 
While, such calibration method requires rather long term and continuous observations, and also the precision data of Earth geopotentials as inputs.
For related missions, to avoid these technical difficulties and to make use of the IS data in time, we suggest here to roll the satellite to give rise to a quick calibration of the intrinsic biases with only the in-orbit measurements as inputs. 

According to Eq. (\ref{eq:ac}) and (\ref{eq:acV}), for the rolling maneuver, we re-write the actuation acceleration measurements as 
\begin{eqnarray}
    a_c^i(t)&=&b^i+k_1^{i\alpha}V_{\alpha}(t) \nonumber\\
&=&\sum_{J=1}^3 a^J_{para,SC}(t) \cos(\Theta^{iJ}(t)+\Theta^{iJ}_0)-G^{ij}(t)d_j + a^i_{para,TM},\nonumber\\
\Rightarrow \ k_1^{i\alpha}V_{\alpha}(t)&=&-b^i+\sum_{J=1}^3 a^J_{para,SC}(t) \cos(\Theta^{iJ}(t)+\Theta^{iJ}_0)-G^{ij}(t)d_j + a^i_{para,TM}.\label{eq:acb} 
\end{eqnarray}
Here $a^J_{para,SC}$ with $(J=X,\ Y,\ Z)$ denotes the components of the non-gravitational accelerations exerted by the satellite in the orbit coordinates system.
$\Theta^{iJ}$ is the angle between the $ith$ axis of the IS system and the $Jth$ axis of the orbit coordinates system.
This rolling modulation will separate the DC and low frequency non-gravitational forces from the intrinsic biases of the IS in the linear acceleration measurements, and could be subtracted or averaged out from the data to suppress their effects on bias estimations.

For practical use, this method benefits when the maneuver time span for each estimation was short, that for short orbital arcs the non-gravitational forces could be treated as varies linearly with time, 
\begin{equation}
  a^J_{para,SC}(t)=a^J t + a^J_0, \label{eq:aJlinear}   
\end{equation}
see Fig. \ref{fig:bias1}(b) in subsection \ref{subsec:bias principle} for illustrations.
The input data sets include the angles $\Theta^{iJ}$, which can be derived by the POD data from GPS or Beidou system and the satellite attitude data from the star trackers, the actuation voltages that are readout by the FEE unit of the IS, and also the scale factors and COM offset been calibrated.
The periodic terms $a^J_{para,SC}(t) \cos(\Theta^{iJ}(t)+\Theta^{iJ}_0)$ on the right-hand side of Eq. (\ref{eq:acb}) can be fitted and subtracted from the IS actuation accelerations. 
Generally, with the in-orbit {  center of mass} adjustment for the satellite platform, the COM offset term $G^{ij}(t)d_j$ could then be ignored in the data fittings.
IF not, the $G^{ij}(t)d_j$ term could also be modeled with the above input data and subtracted from the IS readouts.
Then, the biases can be estimated based on the above observation equation (\ref{eq:acb}).
For Taiji-1, to fulfill the requirement discussed above, the rolling period of the satellite was about $724\ s$, and to test the effectiveness of this method and also to accumulate data segments with better qualities, the entire time span of the rolling maneuver was $1.6\times 10^6\ s$, see Fig. \ref{fig:bias1} for illustrations.

\begin{figure}[h]
\centering
\includegraphics[scale=0.45]{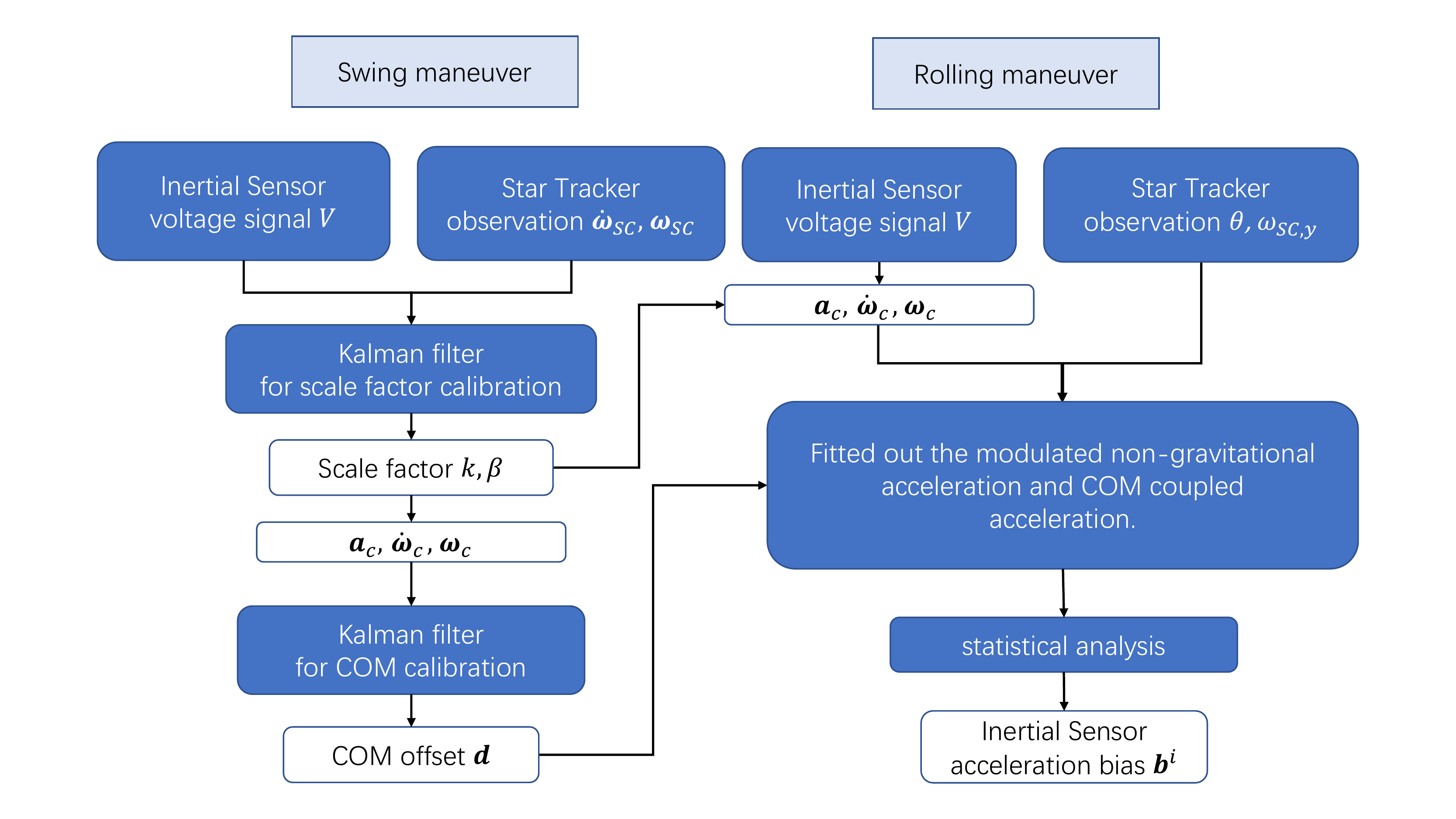}
\caption{The flow chart of the scale factors, COM offset and acceleration bias calibrations.}
\label{fig:flow chart}
\end{figure}

To conclude this section, the complete calibration process of the scale factors, COM offset and IS biases is summarized in Fig. \ref{fig:flow chart}.
%%%%%%%%%%%%%%%%%%%%%%%%%%%%%%%%%%%%%%%%%%

\section{Calibration results \label{RES}}

In 2022 the final extended phase of Taiji-1, still many experiments were planned and performed. 
The swing maneuvers for the IS scale factors and COM offset calibrations were conducted for several trials with slightly different frequencies to obtain a more accurate triangular wave trains of the SC/TM angular velocities or square wave trains of the angular accelerations.
In this work, the scale factors and COM offset calibrations are based on the data of the swing maneuver conducted on 18 MAY 2022.
After the encounter of the satellite with the Earth's shadow, the calibration experiments were continued in { August} 2022.
The rolling maneuver for IS bias calibration was conducted from { August} 2022, lasted for about $1.6 \times 10^6\ s$ to accumulate enough data. 
The detailed data processing procedures and fitting algorithms are expended in the following subsections.

%%%%%%%%%%%%%%%%%%%%%%%%%%%%%%%%%%%%%%%%%%%%%%%%%%%%%%%%

\subsection{Swing maneuver and data preprocessing}

\begin{figure}[htbp]
\centering
\includegraphics[scale=0.35]{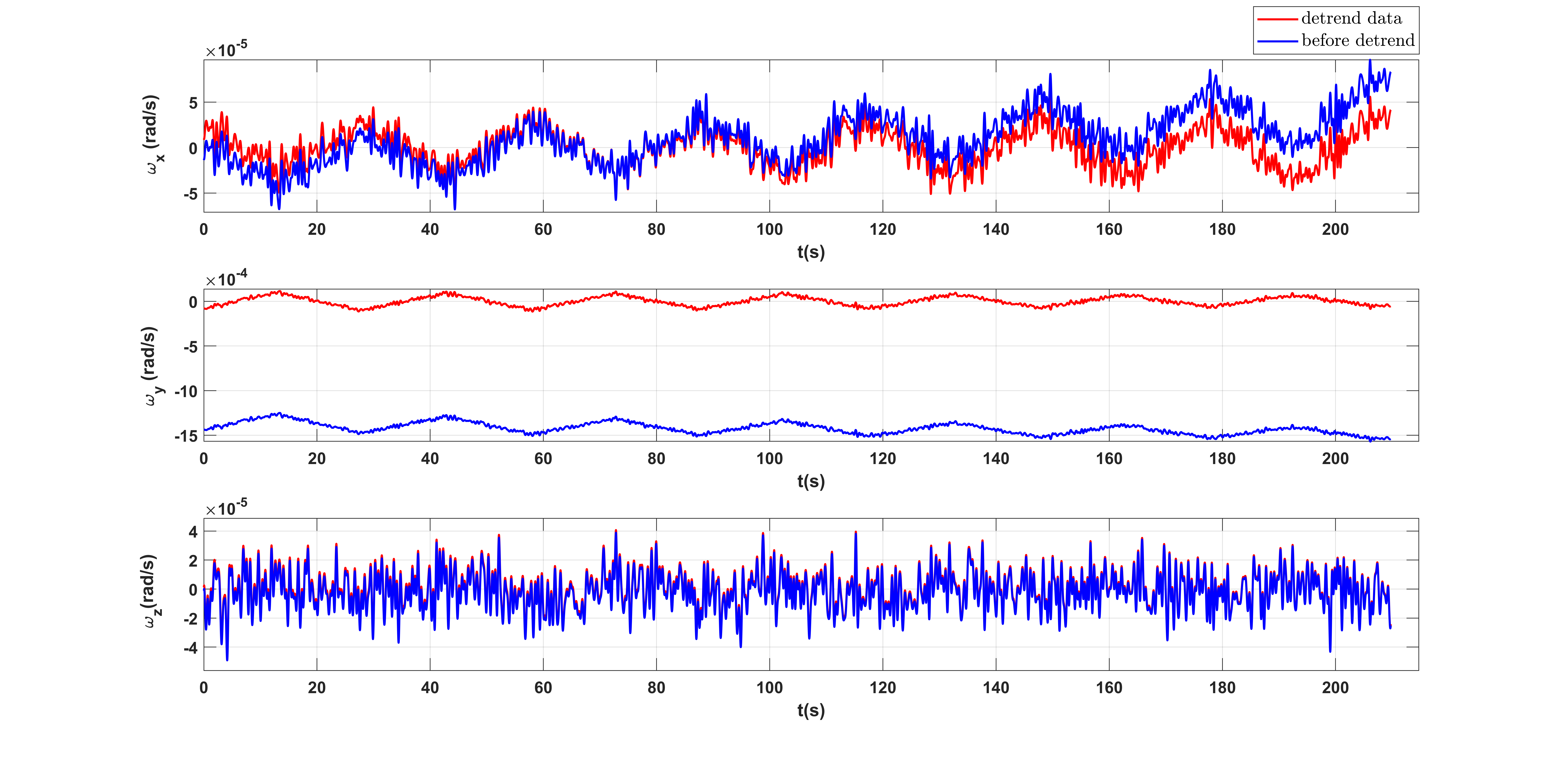}
\caption{The angular velocity of the satellite measured by star tracker. The blue lines denote the data before detrend and the red lines denote the data after detrend.}
\label{fig:pre3}
\end{figure}

\begin{figure}[htbp]
\centering
\subfigure[Satellite angular velocities by star trackers.]{
\begin{minipage}[htb]{1\textwidth}
\includegraphics[scale=0.35]{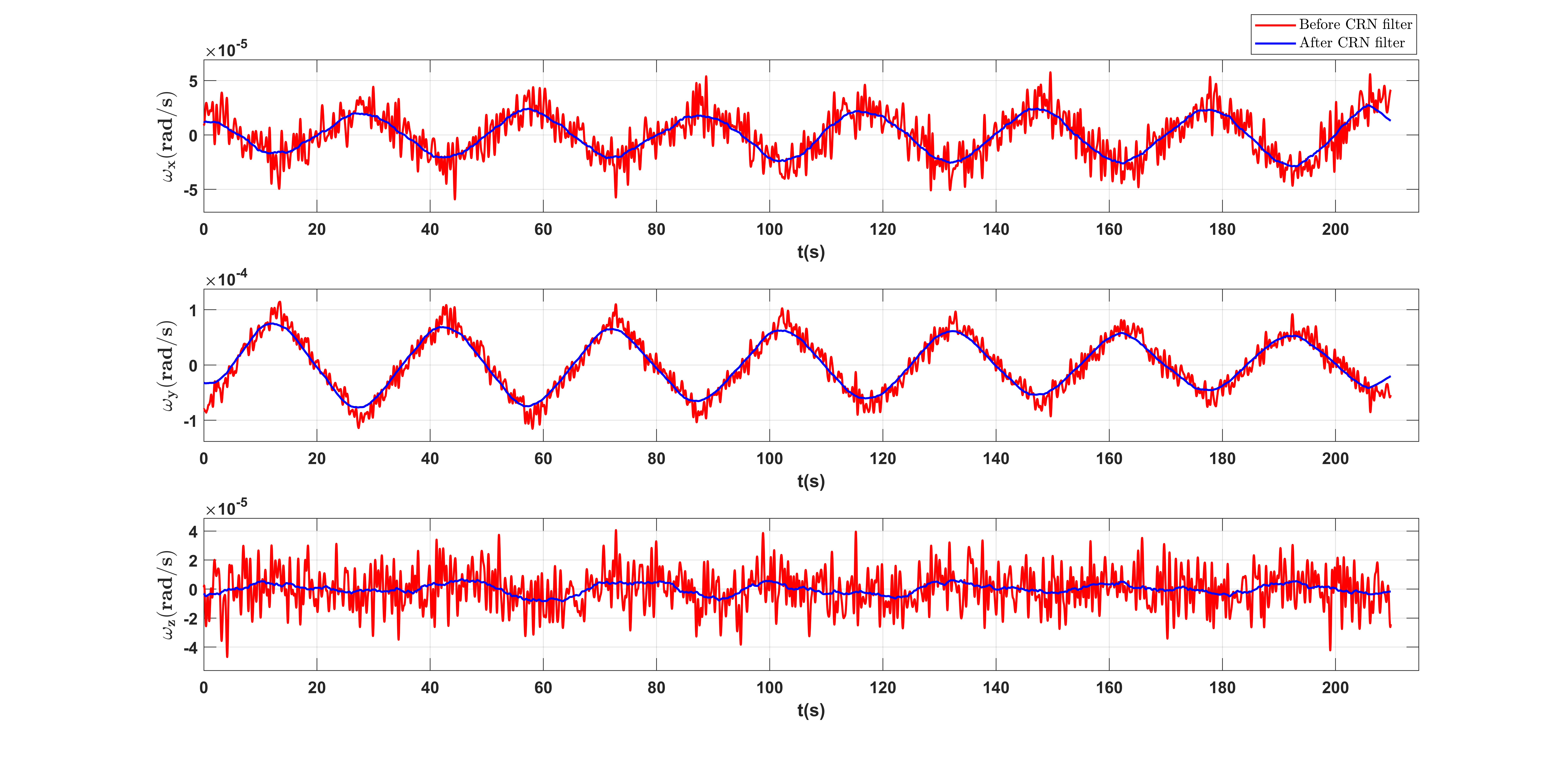}
\end{minipage}
}
\subfigure[Satellite angular accelerations by star trackers before filtered.]{
\begin{minipage}[htb]{1\textwidth}
\includegraphics[scale=0.35]{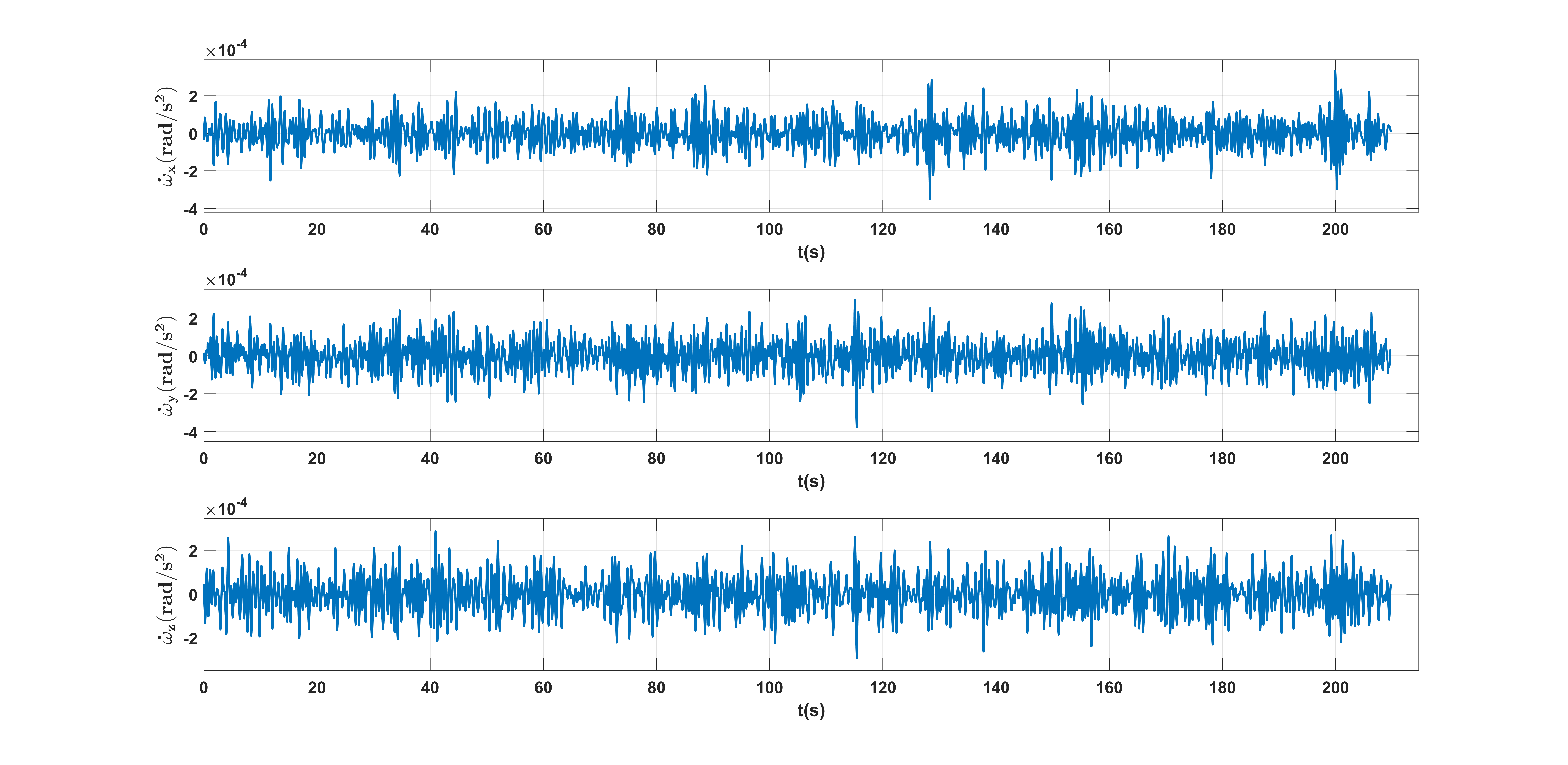}
\end{minipage}
}
\subfigure[Satellite angular accelerations by star trackers after filtered.]{
\begin{minipage}[htb]{1\textwidth}
\includegraphics[scale=0.35]{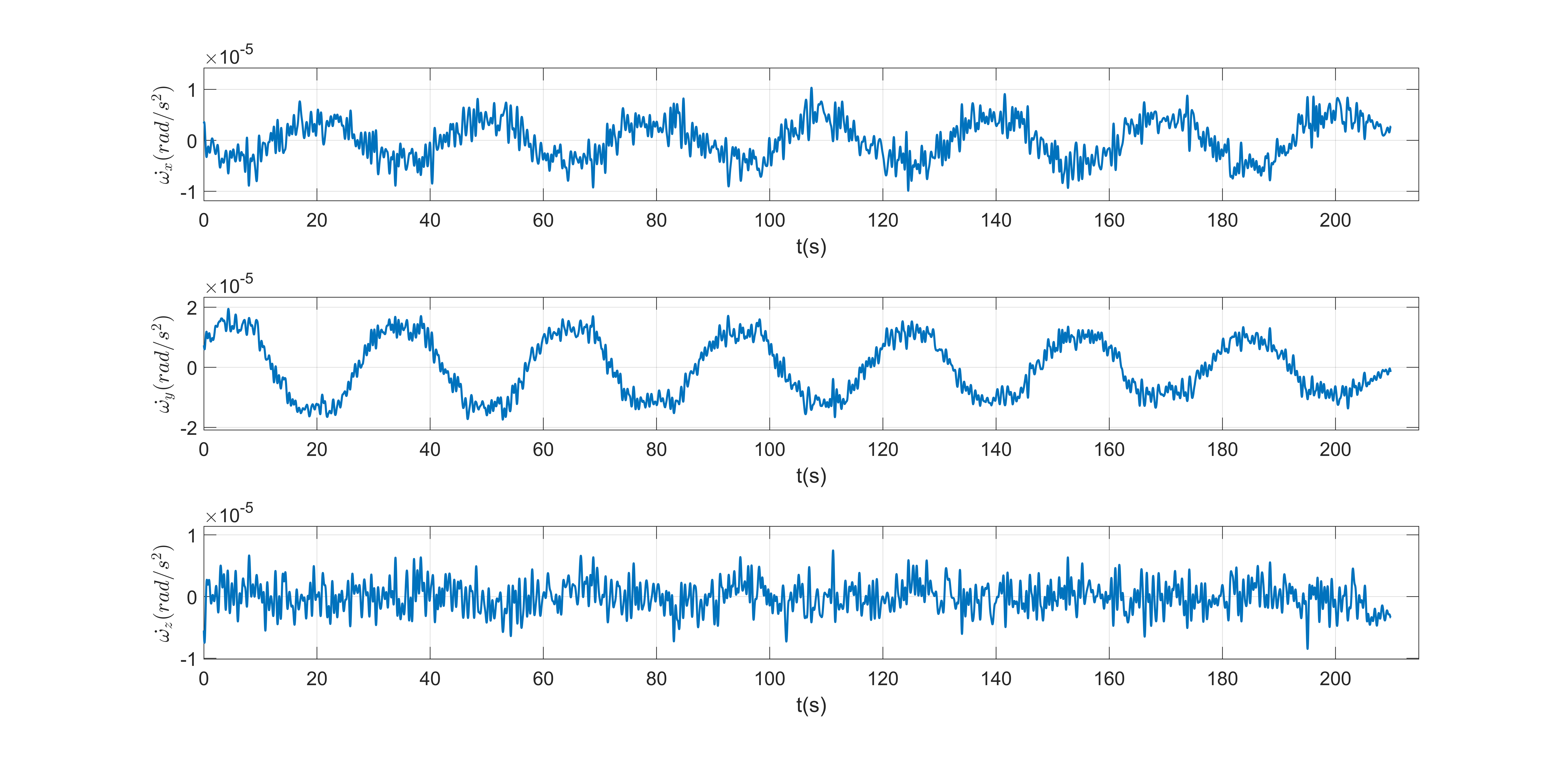}
\end{minipage}
}

\caption{(a) Satellite angular velocities measured by star tracker where Red lines denote the raw readouts and blue lines denote the data filtered by CRN filter. (b) Satellite angular acceleration measured by star tracker. (c) The filtered  satellite angular acceleration filtered by CRN filter.}
\label{fig:pre1}
\end{figure}

\begin{figure}[htbp]
\centering
\includegraphics[scale=0.35]{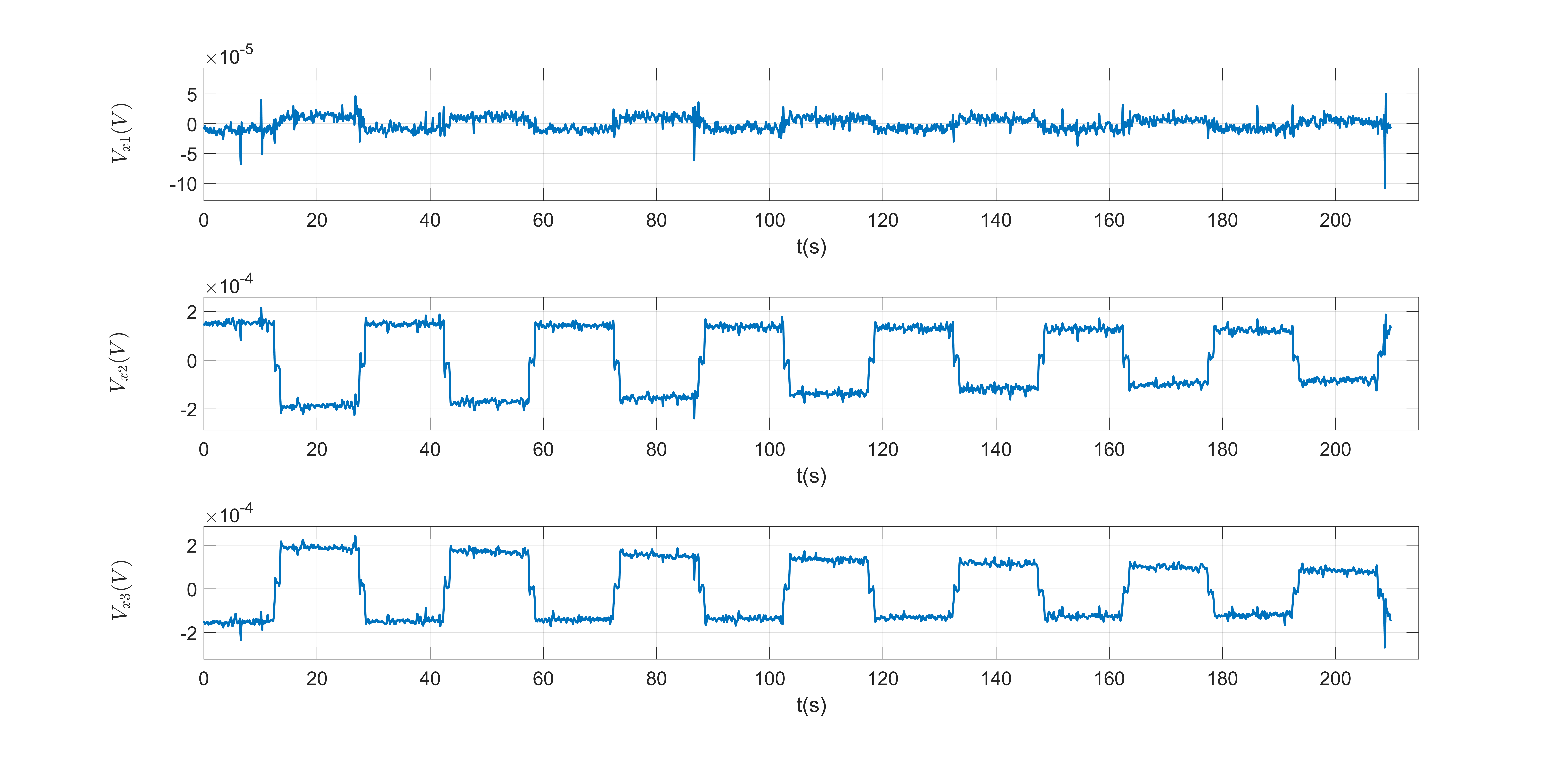}
\caption{IS voltage readouts { along} x-axis during the swing maneuver. }
\label{fig:pre2}
\end{figure}

\begin{figure}[htbp]
\centering
\includegraphics[scale=0.35]{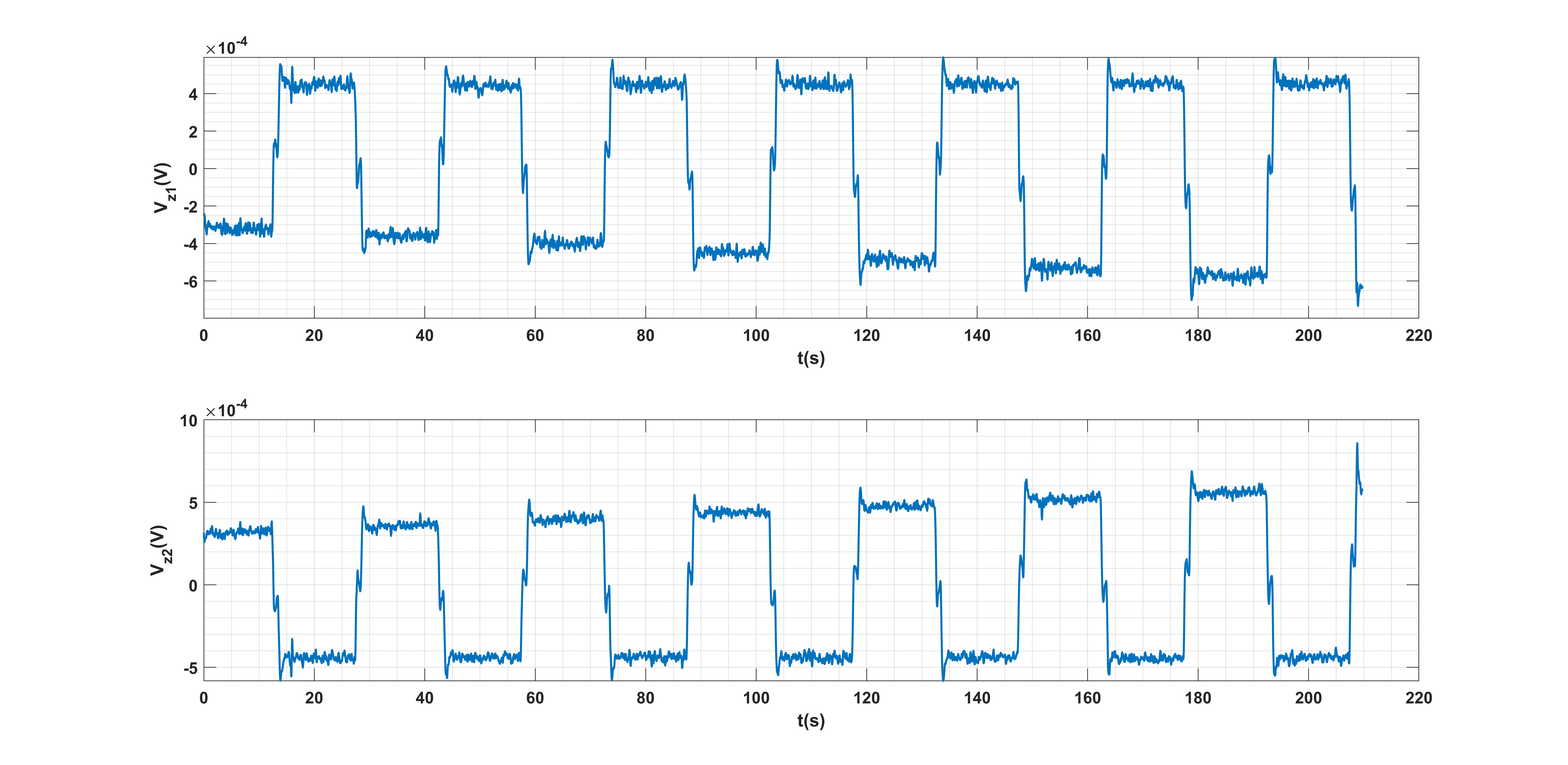}
\caption{IS voltage readouts { along} z-axis during the swing maneuver. }
\label{fig:V_z}
\end{figure}

As discussed in the subsection \ref{subsec:SF&COM principle}, for scale factors and COM offset calibrations, the related data products are the IS actuation voltages readouts, and the satellite attitude data from star trackers. 
The POD data from the Beidou or GPS system is also required to determine the position of the satellite and the local orbital coordinates system.
Then, the satellite attitude variations in the local orbital coordinates system is derived first.
According to the calibration principles, the Fourier components of the swing frequency and its harmonics in the data are used to fit the corresponding parameters. 
To reduce possible interferences from the high frequency noises (mainly comes from the FEE unit) and long time drifts, the actuation voltages and attitude data are detrended, and then smoothed with low path filters.
The low-path filter used here is the CRN filter (a classical digital filter characterized by an N-th order self Convolution of Rectangular time-domain window function) with the cut-off frequency 0.1 Hz. 
One sees Fig. (\ref{fig:pre3}) - (\ref{fig:V_z})  for the illustrations of the attitude and IS voltages data.

From these figures, one sees that the projections of the swings along the z-axis (IS frame) is rather obscure. 
This is acceptable as it is mentioned in the previous section that for Taiji-1's IS the scale factor $k^y$ can not be calibrated by this method, and for the COM offset calibrations only swings along two different orthogonal axes are needed because of the redundancy.
The outliers or spikes, for example the $V_{x1}$ data, are kept in the data, since no evident correspondence between such possible ``data anomalies'' with any instruments or payload events are found.

%%%%%%%%%%%%%%%%%%%%%%%%%%%%%%%%%%%%%%%%%%%%%%%%%%%%%%%%%%

\subsection{Results of scale factor calibrations} \label{sec4.2}

{ As discussed in subsection \ref{subsec:SF&COM principle}, our approach is based on the relations between the linear and angular scale factors showed in Eqs. (\ref{e7}) - (\ref{e9}). These equations depend only on the parameters of the TM, which can be measured precisely before launch and hardly change during the mission lifetime. Therefore,with the data from the IS and star trackers, one calibrates the angular scale factors first, and then wtih Eqs. (\ref{e7}) - (\ref{e9})  the estimations of the linear scale factors could be obtained. }

After the preparations of the IS voltages and satellite attitude data sets, base on the method discussed in subsection \ref{subsec:SF&COM principle}, the angular scale factors of Taiji-1's IS ([$\beta^x, \ \beta^y, \ \beta^z$] with $\beta^y=\beta^z$) can be estimated by means of the following algorithms of Kalman filters 
\begin{align}
x_{n+1} &= \Phi_n x_n + \Gamma_nu_n+\Upsilon_n w_n \label{eqn:kal1},\\ 
\hat{y}_n &= H_nx_n +\epsilon_n \label{eqn:kal2}. 
\end{align}
{ For sequential least square algorithm, we found that the estimations were hard to converge and oscillating around the mean values. The Kalman filter algorithm is adopted here since it is further adjusted in the Kalman gain matrix, which helps to improve the performance of the estimator.}
The above equations are the standard form of the discrete-time linear Kalman filter, and $\Phi$ is the state transition matrix, $\Gamma$ the gain of the input $u$ in the prediction/propagation equation, and $\Upsilon$ the gain of the noise $w$ of the dynamics of the estimator $x$.
The number $n$ represents the sample or the step number and the ``$\hat{\ }$'' labels the estimation values. 
Here, the estimator $x$, input $y$ and model matrix $H$ are defined as
\begin{align}
x &= [\beta^x, \quad \beta^y]^T, \label{eqn:sca1} \\ 
y &= [\dot{\omega}_x, \quad \dot{\omega}_y]^T, \label{eqn:sca2}
\end{align}
\begin{equation}
H =\left(
\begin{array}{cc}
    V_{z1}-V_{z2} & 0  \\
    0 & V_{x2}-V_{x3}.   
\label{eqn:sca3}
\end{array}
\right).
\end{equation} 
For $\beta^x$ and $\beta^y$ are constant parameters without dynamic feature, we have
\begin{align*}
\Phi_k &= 1 , \\ 
\Gamma_k &= \Upsilon_k = 0 . 
\end{align*}
In this case, to minimize the measurement noise $\epsilon$, the Kalman filter equations can be re-written as
\begin{align}
\hat{x}_{n+1} &= x_n + K_n(\hat{y}_n - H_n\hat{x}_n), \label{eqn:kal5}\\ 
K_n &= P_n{H_n}^T[(H_nP_n{H_n}^T)+R_n ]^{-1},\label{eqn:kal6}\\
P_{n+1} &= [I-K_nH_n]P_n\label{eqn:kal7},
\end{align}
where the error covariance matrix $P$ and variance matrix $R$ of the measurement noise are defined as
\begin{align}
P_n &= E\{\hat{x}_n\hat{x}^T_n\} \label{eqn:kal8}, \\
R_n &= E\{\hat{\epsilon}_n\hat{\epsilon}^T_n\} \label{eqn:kal8}. 
\end{align}
{ 
The initial values $P_0 = E\{\hat{x}_0\hat{x}^T_0\}$ and $R_0 = E\{\hat{\epsilon}_0\hat{\epsilon}^T_0\}$ are obtained from a prior calibration model using the sequential least square method.
% To obtained the initial value of $x$ and $\epsilon$, a linear squential estimation was applied on scale factor calibraiton as the prior model, which the corresponding result of this estimation will be regarded as the initial values of $P_0$ and $R_0$ in Kalman filter.
}

\begin{figure}[htbp]
\centering
\includegraphics[scale=0.36]{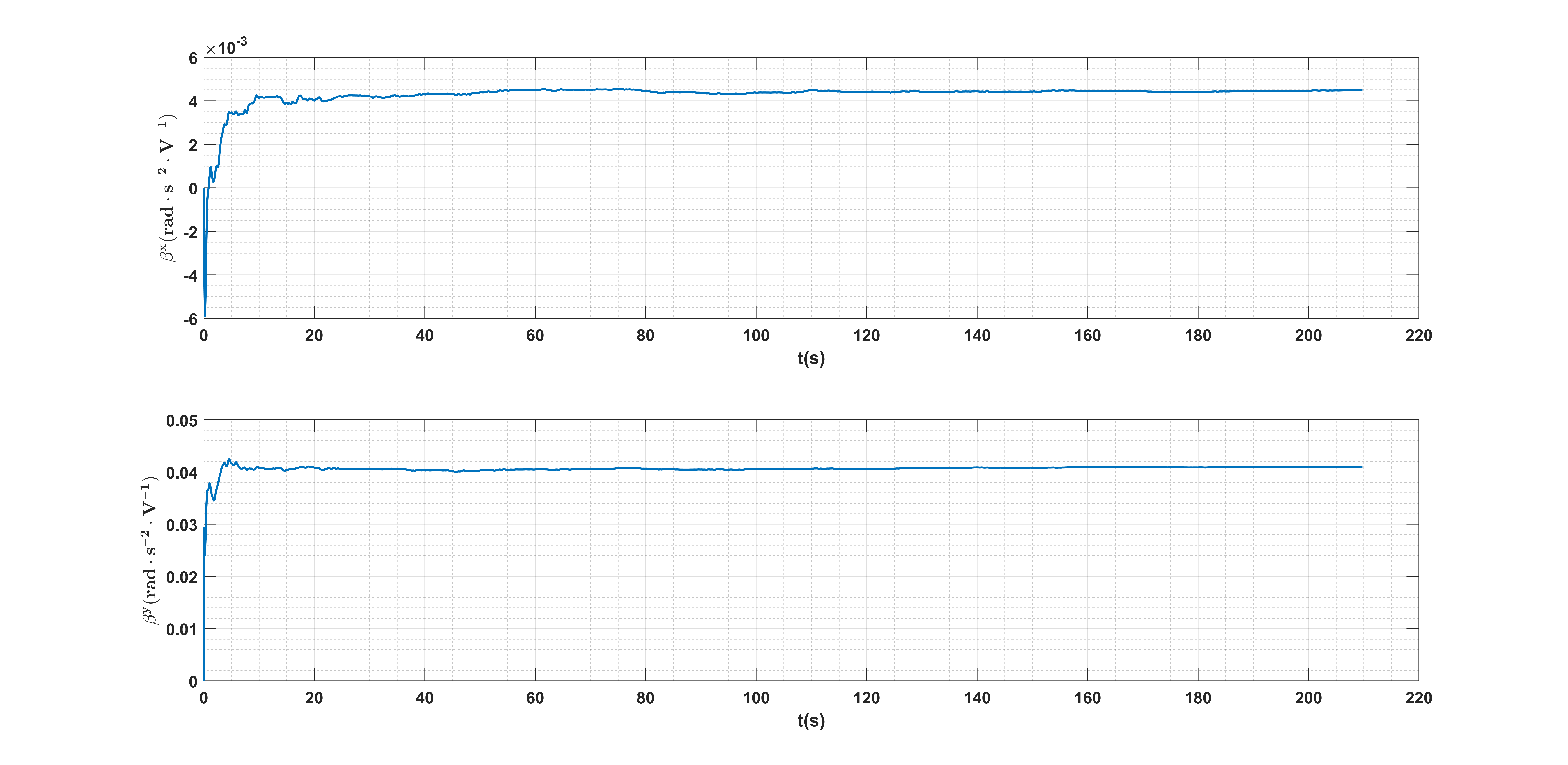}
\caption{Convergences of the angular scale factors $\beta_x$ (top) and $\beta_y$ (bottom) in the estimations.}
\label{fig:scal1}
\end{figure}

\begin{figure}[htbp]
\centering
\includegraphics[scale=0.36]{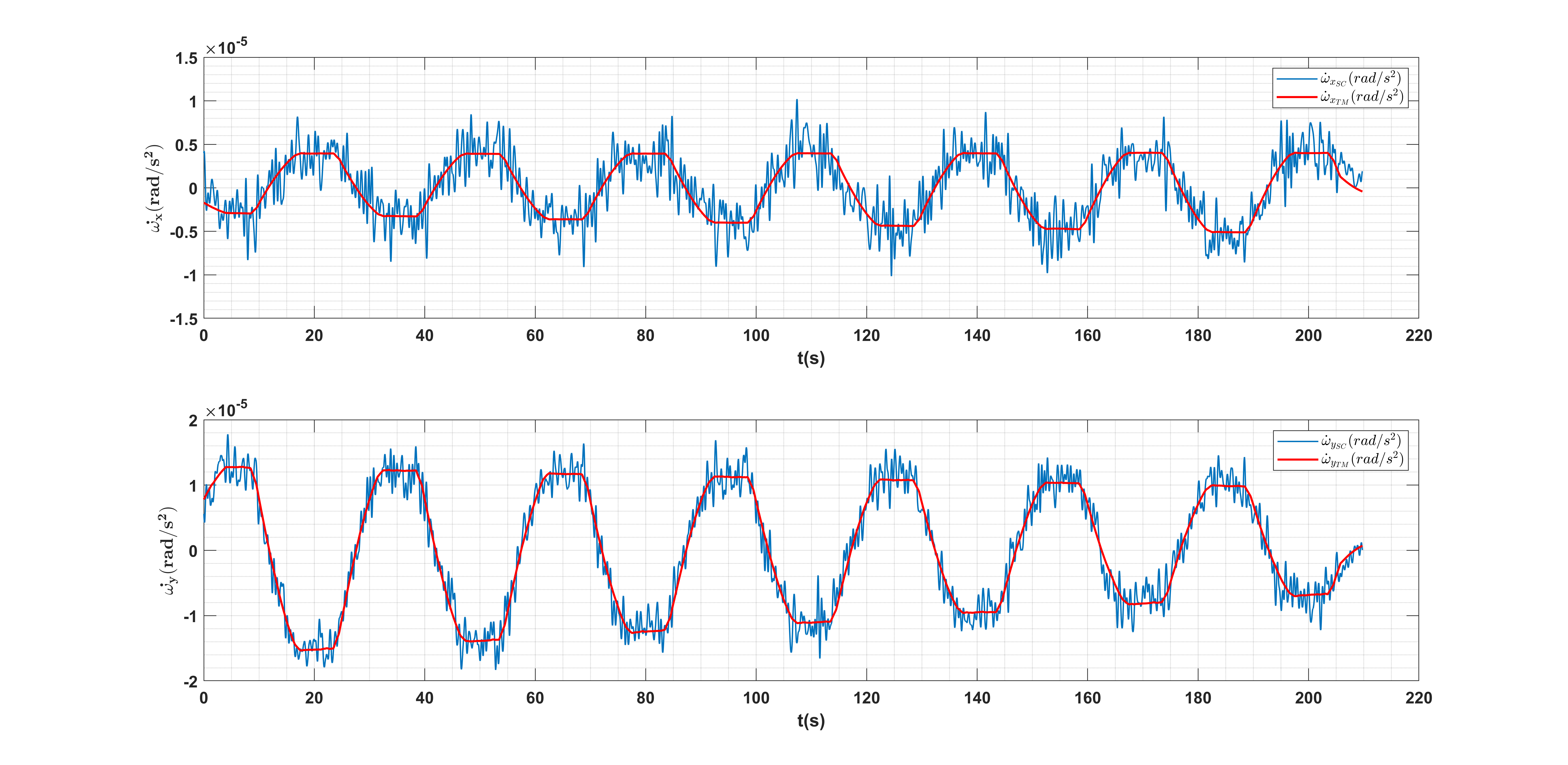}
\caption{The comparisons between the angular accelerations from the star trackers (blue lines) and the modeled actuation voltages with calibrated scale factors (red lines) during the swing maneuver.}
\label{fig:scal2}
\end{figure}

\begin{figure}[htbp]
\centering
\includegraphics[scale=0.36]{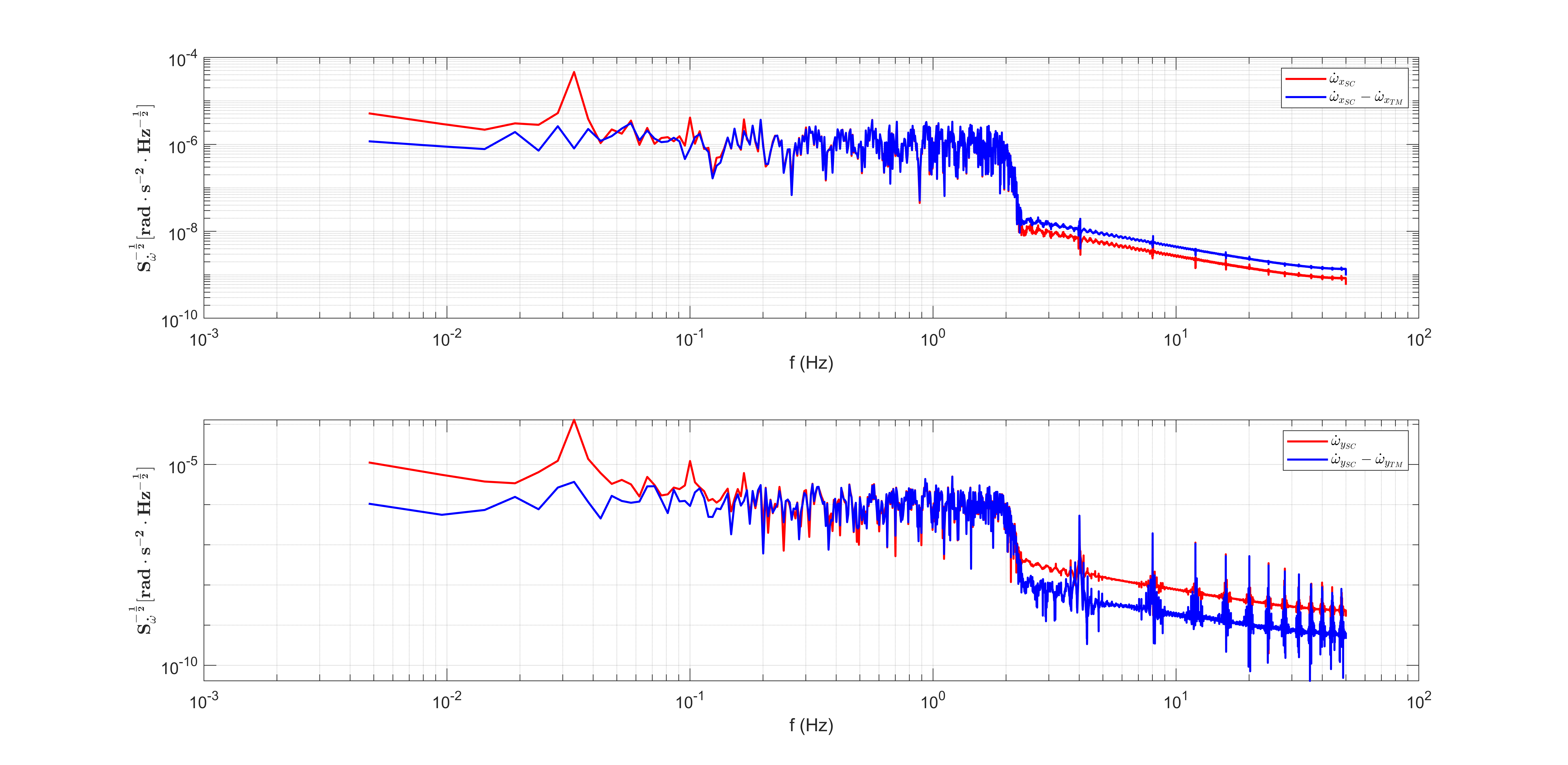}
\caption{The blue lines are the ASD curves of the difference between the angular accelerations from the star trackers and the modeled actuation voltages with calibrated scale factors during the swing maneuver. The red lines are the ASD of the angular accelerations from the star trackers.}
\label{fig:dp_asd}
\end{figure}

With this algorithm and given the prepared data from the swing maneuver, the angular scale factors $\beta^x$ and $\beta^y$ convergent rather fast to their estimated values, see Fig. \ref{fig:scal1} for illustrations.  
{ In this work, the accuracies of determining the parameters in Kalman filters are adopted as the standard deviations of the estimated values after they had converged.}
The comparisons between the angular accelerations from the star trackers and the modeled actuation voltages with calibrated scale factors are shown in Fig. \ref{fig:scal2}, and their residuals are shown in %Fig. \ref{fig:dp_t} and 
Fig. \ref{fig:dp_asd}.

%\begin{figure}[htbp]
%\centering
%\includegraphics[scale=0.36]{picture/dp_and_p_t_scl.png}
%\caption{The difference between the angular accelerations from the star tracks and the modeled actuation voltages with calibrated scale factors during the swing maneuver.}
%\label{fig:dp_t}
%\end{figure}

At last, with the relations Eq.(\ref{e7}) and (\ref{e8}), the complete scale factors  calibrated with this method together with their estimation errors are listed in Tab. \ref{tabel2}.
\begin{table}[htbp]
\begin{center}
\caption{Scale factors in-orbit calibration results, compared with their nominal values and ground-based calibration values.}
\label{tabel2}
\begin{tabular}{c|c|c|c}
\hline
Scale factor & Nominal values & Calibrated values & Calibrated values \\
& & On-ground & In-orbit \\
\hline
$k^x(m/(s^2\cdot V))$& $5.41\times{10^{-4}}$&null&$5.43\times{10^{-4}}\pm 3.73\times{10^{-6}}$\\
$k^y(m/(s^2\cdot V))$& $3.00\times{10^{-4}}$&$2.7\times{10^{-4}}$&null\\
$k^z(m/(s^2\cdot V))$& $1.58\times{10^{-4}}$&$1.45\times{10^{-4}}$&$1.14\times{10^{-4}}\pm 2.67\times{10^{-7}}$\\
$\beta^x(rad/(s^2\cdot V))$&$5.93\times{10^{-3}} $&$6.21\times{10^{-3}}$&$4.42\times{10^{-3}}\pm 2.81\times{10^{-5}} $\\
$\beta^y(rad/(s^2\cdot V))$&$3.82\times{10^{-2}}$&null&$4.08\times{10^{-2}}\pm 9.57\times{10^{-5}} $\\
$\beta^z(rad/(s^2\cdot V))$&$3.82\times{10^{-2}}$&null&$4.08\times{10^{-2}}\pm 9.57\times{10^{-5}} $\\
\hline
\end{tabular}
\end{center}
\end{table}

%%%%%%%%%%%%%%%%%%%%%%%%%%%%%%%%%%%%%%%%%%%%%%%%%%%%%%%%%%%
\subsection{Results of COM calibration}
With the re-calibrated scale factors, the actuation voltage signals of the IS are transformed to the linear compensation acceleration $a^i_{c}$ and angular compensation acceleration $\dot{\omega}_{c}^i$. 
Compared with the star tracker data, the attitude variation signals of the TM (or the satellite) measured by the IS have a better SNR. 
Therefore, based on the discussion in the previous section, that during the swing maneuver the IS was in its normal ACC mode and the TM followed tightly the rotations of the satellite platform, we will use the attitude variation data $\dot{\omega}_{TM}$ measured by the IS instead of $\dot{\omega}_{SC}$ derived by the star trackers data in the following COM offset calibrations.

The estimation algorithm based on Kalman filters is the same as that for the scale factors estimations in the previous subsection.
The COM offset vector $\vec{d}$ is an unknown constant vector without dynamical features, therefore the equations of parameter estimations can be re-written in the form of Eq. (\ref{eqn:kal5}) - (\ref{eqn:kal8}). 
% state transition matrix $\Phi(t,t_0)$ in Eq. (\ref{eqn:kal1}) would be unit matrix as well. Besides that, the linear term related to non-gravitational force and gravity gradient in Eq. (\ref{e9}) is eliminated after preprocessing in advance, which means the estimator and relating parameters in COM offset could be simplified as:
Here, for the case of COM offset calibration, the estimator $x$, input $y$ and model $H$ are 
\begin{align*}
x&=[d_x,\quad d_y,\quad d_z]^T,\\
y&= [a^x,\quad a^y,\quad a^z]^T,\\
H&= \mathbf{A}(\omega_i,\dot{\omega}_i),
\end{align*}
{ 
where $A(\omega,\dot{\omega})$ are defined in Eq.(\ref{eqn_A}}). 

\begin{figure}[H]
\centering
\includegraphics[scale=0.37]{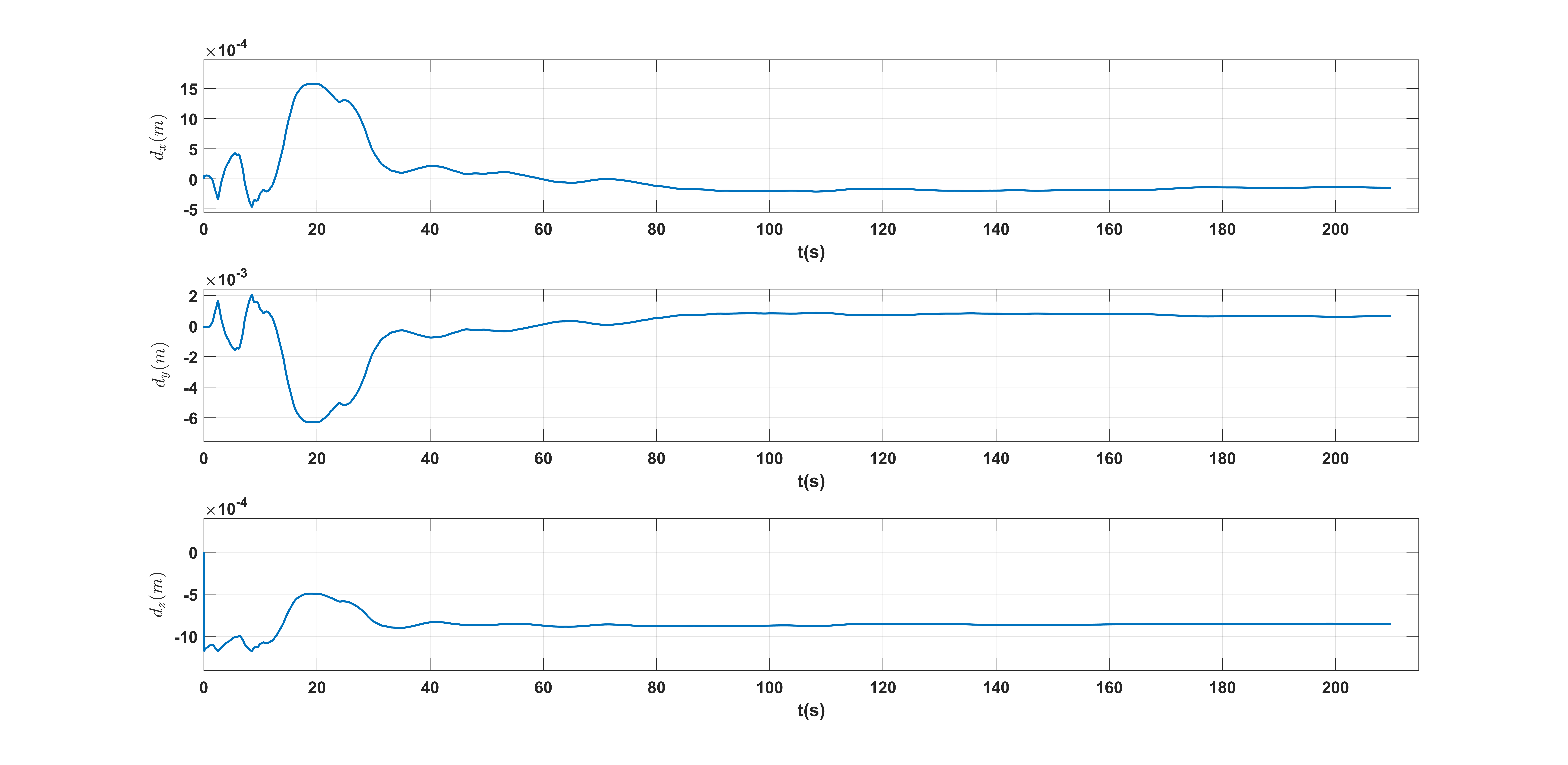}
\caption{Convergence of the COM offset $d_i$ in the estimations.}
\label{fig:com_cal1}
\end{figure}

\begin{figure}[H]
\centering
\includegraphics[scale=0.37]{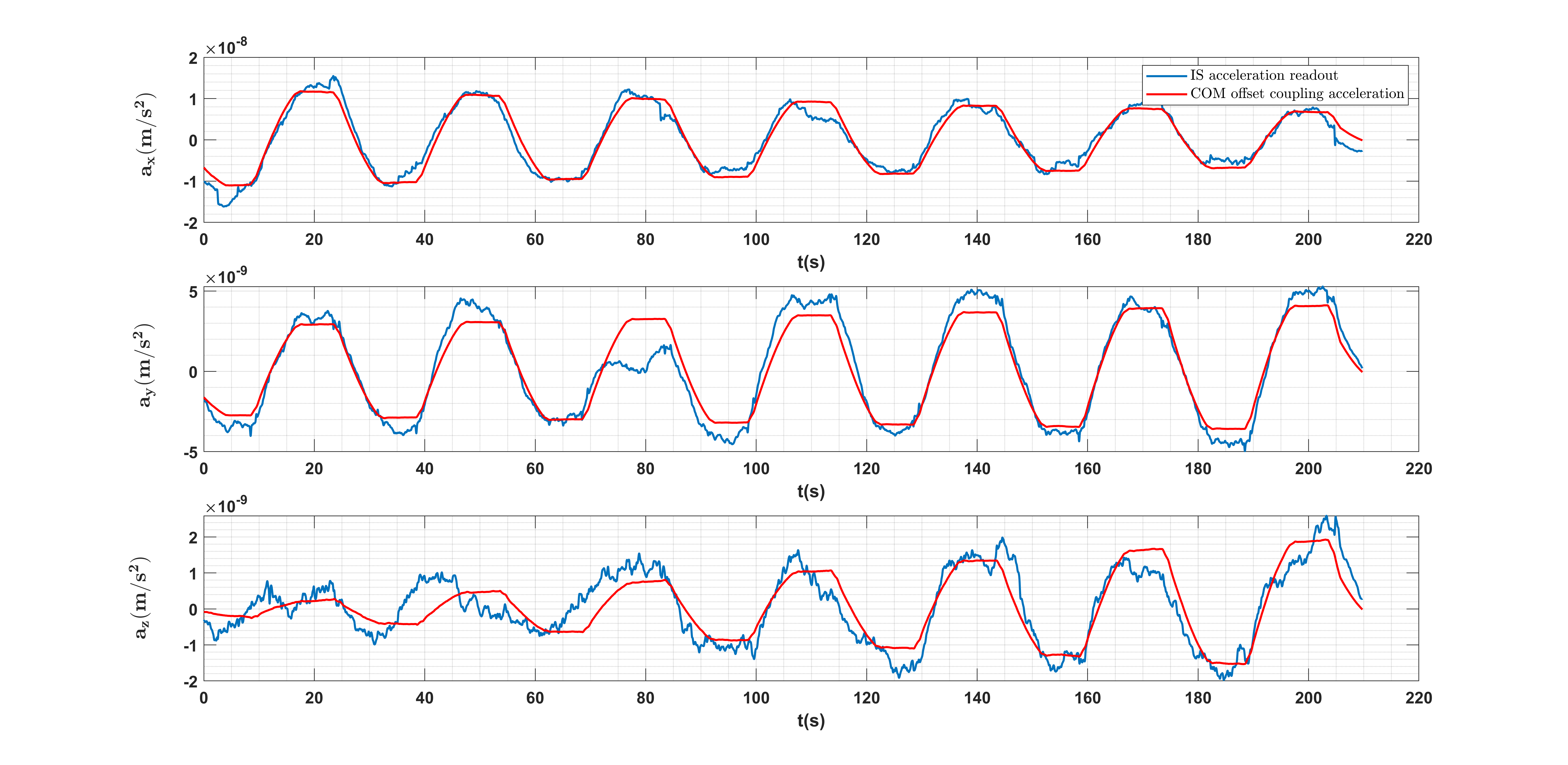}
\caption{Comparison between the filtered TM linear acceleration measurements (blue lines) and the modeled inertial accelerations with calibrated COM offset (red lines) during the swing maneuver.}
\label{fig:com_cal2}
\end{figure}

\begin{figure}[H]
\centering
\includegraphics[scale=0.4]{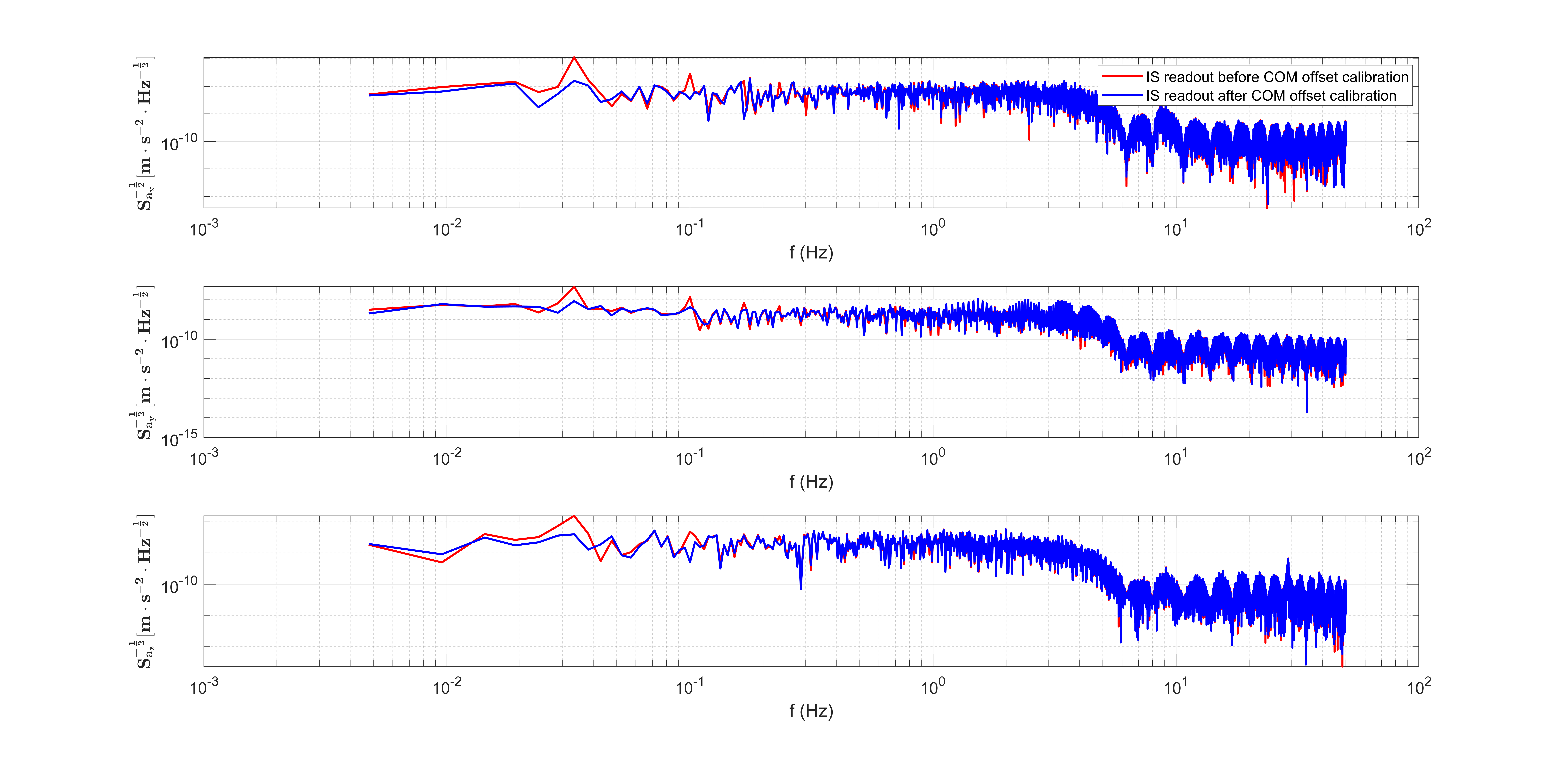}
\caption{
Comparisons of the ASD curves of the TM linear accelerations before (red lines) and after (blue lines) the COM corrections.}
\label{fig.8}
\end{figure}

The convergences of the COM offset components are shown in Fig. \ref{fig:com_cal1}.
The comparisons between the measured linear acceleration (after pre-processing) and the modeled inertial accelerations with the re-calibrated values $d^i$ are shown in Fig. \ref{fig:com_cal2}.
The re-calibrated COM offset vector is suggested to the Taiji-1 science team, and in practical use the inertial accelerations are to be modeled given the attitude data and COM offset values, and be subtracted from the linear acceleration data (COM correction).   
The ASD curves of the TM's linear accelerations before and after the COM corrections during the swing maneuver can be found in Fig. \ref{fig.8}, and with the estimated values $d_i$ the peaks of the inertial accelerations  are successfully identified and removed.
Finally, the calibrated COM offset values and their estimation errors are listed in Tab. \ref{table3}.
After nearly three years operation, the magnitudes of Taiji-1 COM offset are still within the order of $10^2\ \mu m$, while its long-term changes are also evident with respect to its nominal values.
\begin{table}[H]
\begin{center}
\caption{COM offset calibration results for Taiji-1' IS system.}
\label{table3}
\begin{tabular}{c|c|c}
\hline
 COM offset & Calibrated value (\textmu m)&Error (\textmu m)\\
\hline
$d_x$& $-140.02$&$\pm5.02$\\
$d_y$& $627.75$&$\pm16.25$\\
$d_z$& $-896.46$&$\pm1.16$\\
\hline
\end{tabular}
\end{center}
\end{table}

\subsection{Results of bias calibration}

\begin{figure}[htbp]
\centering
\subfigure[Satellite attitude evolution durring the rolling maneuver.]{
\begin{minipage}[htb]{1\textwidth}
\includegraphics[scale=0.32]{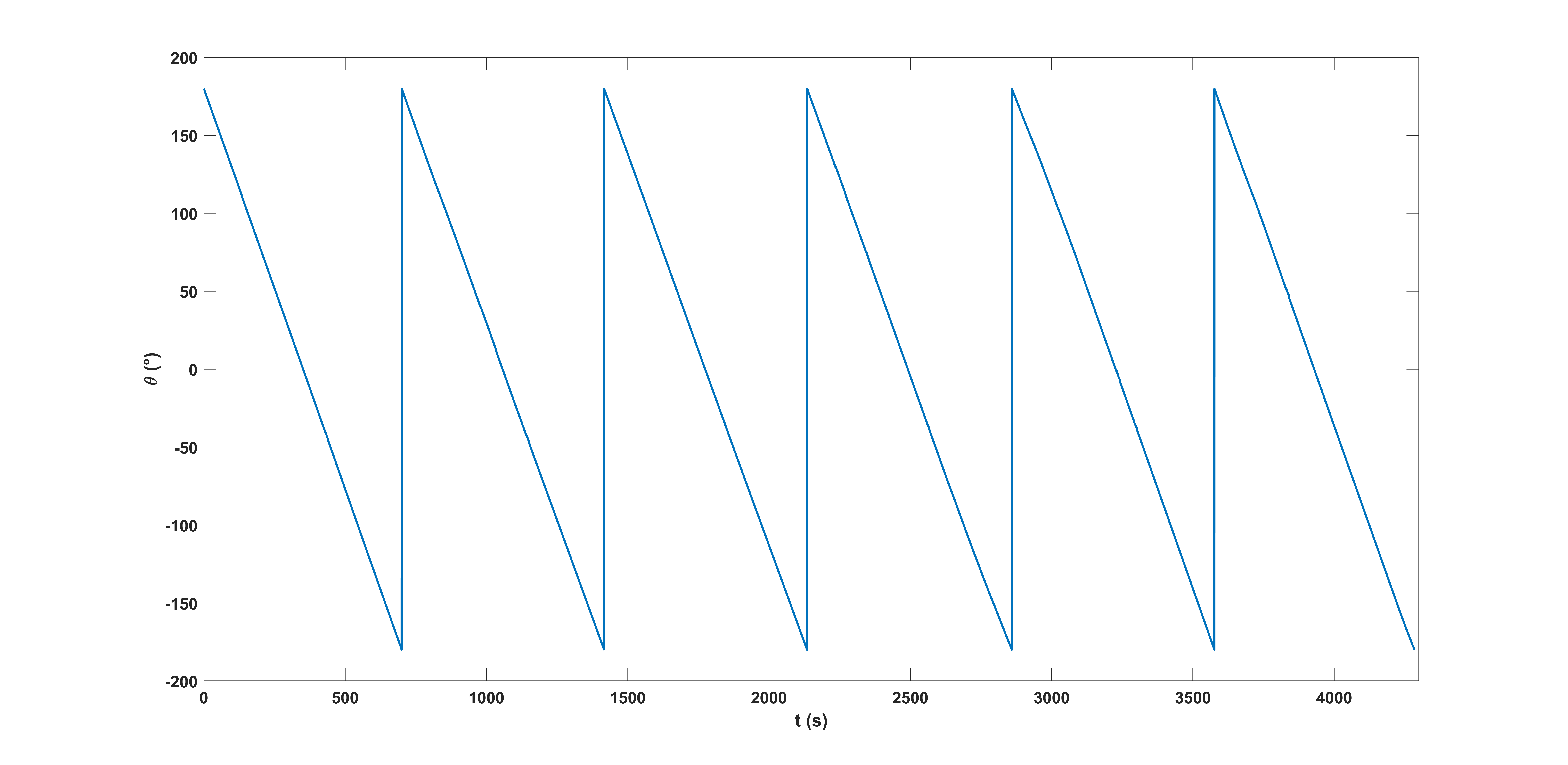}
\end{minipage}
}
\subfigure[TM linear accelerations that modulated by the rolling maneuver.]{
\begin{minipage}[htb]{1\textwidth}
\includegraphics[scale=0.33]{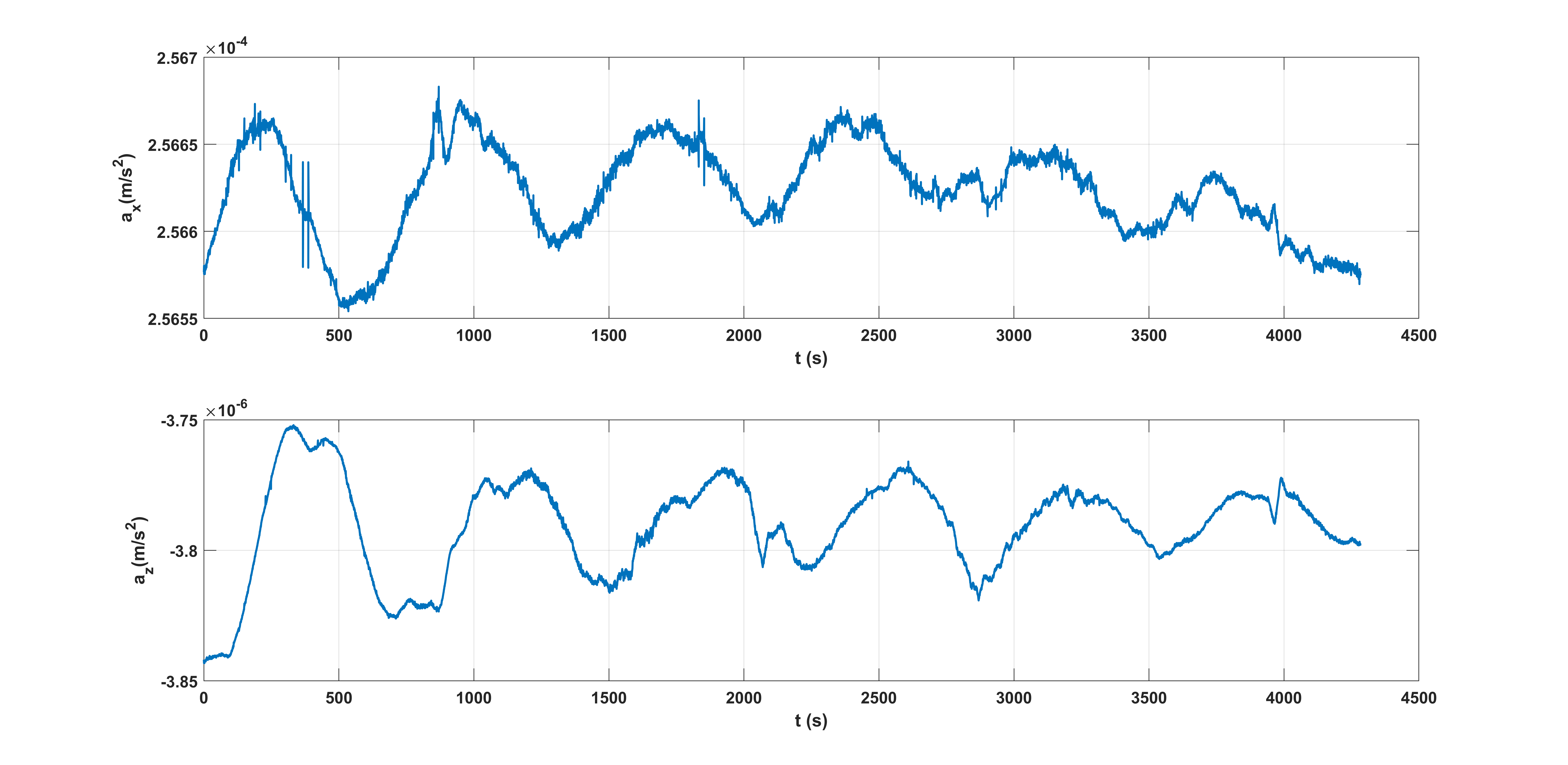}
\end{minipage}
}
\caption{(a). Attitude evolution of the satellite { along} y-axis during the rolling maneuver. (b). The TM linear accelerations along the x and z axes during the rolling maneuver.}
\label{fig:bias1}
\end{figure}

\begin{figure}
    \centering
    \includegraphics[scale=0.35]{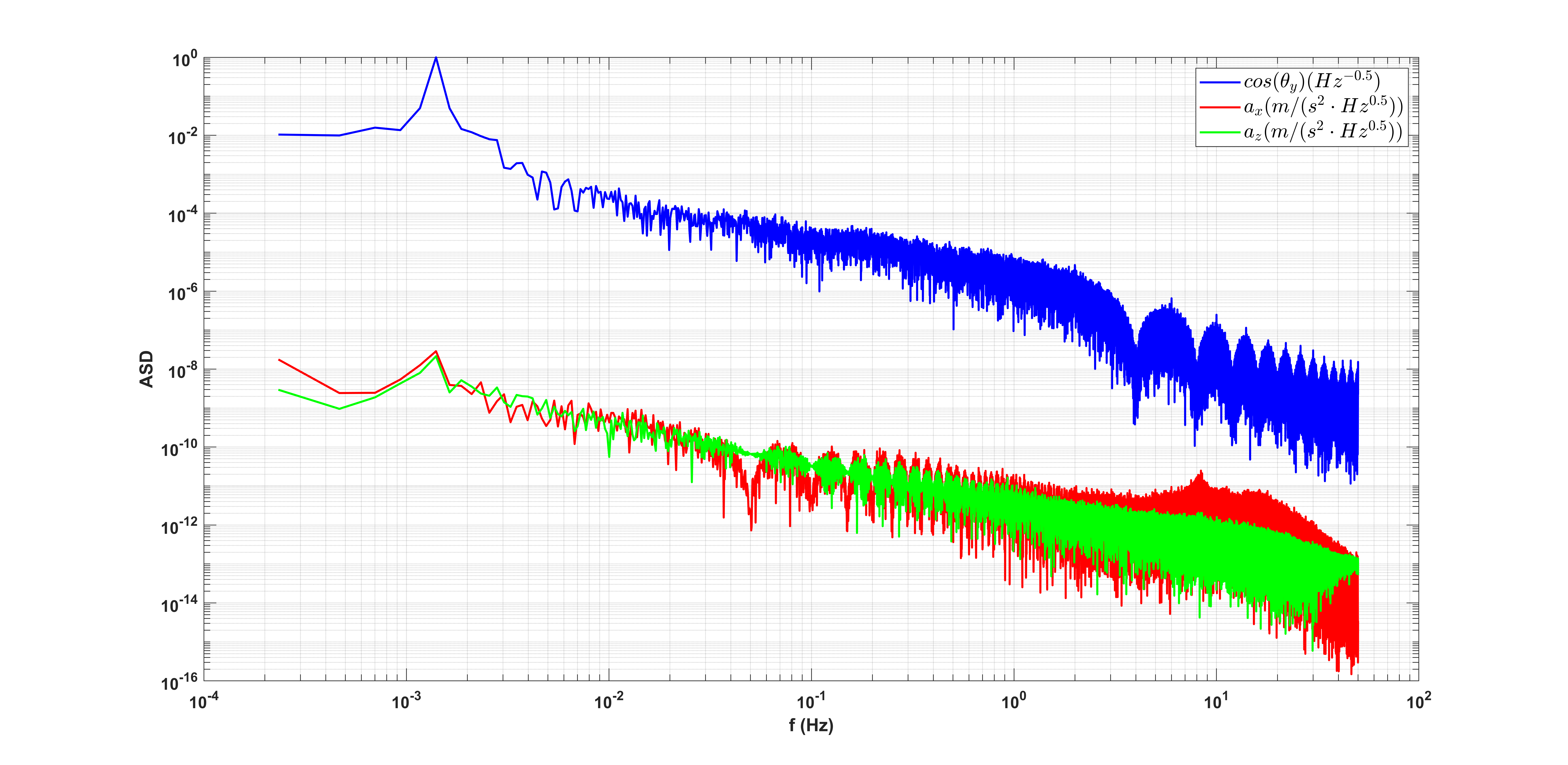}
    \caption{ASD curves of the attitude variation along y-axis and the TM linear acceleration measurements in the x and z axes during the rolling maneuver.}
    \label{fig:asd rolling}
\end{figure}

Based on the principle of bias calibration discussed in Sec. \ref{subsec:bias principle}, the Taiji-1 satellite performed a long time and uniform rolling maneuver { along} y-axis (IS frame) in { August} 2022, which lasted for about $1.6 \times 10^6 \ s$ ($\sim 19\ days$).
The rolling periods were short,  $\sim 724\ s$, that a data segment about $2000 \sim 3000 \ s$ long would contain several rolling periods and the linear approximations of the non-gravitational forces in Eq. (\ref{eq:aJlinear}) could be applied. 
A segment of the time series data of the satellite attitude evolution $\theta(t)$ { along} y-axis is shown in Fig. \ref{fig:bias1}(a), and the linear accelerations
modulated by the rollings in the x-axis and z-axis are shown in Fig. \ref{fig:bias1}(b).
Their ASD curves can be found in Fig. \ref{fig:asd rolling}.
One can see that in  Fig. \ref{fig:bias1}(b) the amplitudes evolutions of the oscillating signals in the TM linear acceleration measurements agree well with the linear model from Eq. (\ref{eq:aJlinear}),
\[
\sum_{J=1}^3 a^J_{para,SC}(t) \cos(\Theta^{iJ}(t)+\Theta^{iJ}_0)=\sum_{J=1}^3 [a^J t\cos(\Theta^{iJ}(t)+\Theta^{iJ}_0)+a^J_0 \cos(\Theta^{iJ}(t)+\Theta^{iJ}_0)].
\]
Other long-term trends due to orbital evolutions, see the linear acceleration in x-axis in Fig \ref{fig:bias1}(b) as an example, can be fitted out with quadratic or cubic polynomial fitting methods.

The estimation procedure for the intrinsic biases of Taiji-1's IS can be summarized as the following five steps: 1) preprocessing including data segmentation and quality check, 2) fitting out trends in the TM linear accelerations due to orbital evolutions, 3) fitting out the oscillating non-gravitational signals and the COM offset coupled signals, 4) estimations of the biases, 5) statistical analysis of the estimations for each data segment.
In preprocessing, the entire data set of the rolling maneuver is divided into short data segments with only a few rolling periods.
The data quality of each segment were checked, and we choose only the data segments that have a more uniform rolling rate. 
About 300 data segments with different lengths are used in the bias calibrations, and the final estimations, see Tab. \ref{table4}, are obtained from the statistical analysis of the calibration results for each segment. 
\begin{table}[htbp]
\begin{center}
\caption{Intrinsic biases in-orbit calibration results for Taiji-1's IS.}
\label{table4}
\begin{tabular}{c|c|c}
\hline
Intrinsic acceleration bias& Calibrated value  $(m/s^2)$& Error $(m/s^2)$\\
\hline
$a_x$& $-2.5840\times{10^{-4}}$&$\pm1.0173\times{10^{-8}}$\\
$a_y$& $1.9488\times{10^{-5}}$&$\pm2.6347\times{10^{-8}}$\\
$a_z$& $2.9887\times{10^{-6}}$&$ \pm 2.9443\times{10^{-9}}$\\
\hline
\end{tabular}
\end{center}
\end{table}

% According to Table.\ref{table4}, both the relative error level of calibration value alone x-axis and z-axis are less than 0.1\% with the help of satellite flip maneuver while the error alone y-axis is about 0.14\%. 

\subsection{IS performance evaluation \label{subsection:performance}}

With all the necessary operating parameters been calibrated in this work,  the performance of the Taiji-1's IS system should be re-estimated with the updated parameters.
In previous works \cite{yue2021china,wang2021development}, the performance or resolution level was obtained based on the y-axis measurements, since the projections of the non-gravitational forces, such as air drags and etc., in the orbital normal direction are small in comparing with those in the flight direction.
But, as discussed in the previous sections, the scale factors $k^y$ can not be calibrated with our new method, therefore the performance of the other sensitive axis, that the z-axis along the flight direction, is re-estimated and discussed here. 
The original estimation of the noise along this axis can be found in Fig. 16(a) from \cite{wang2021development} and in Fig. 2(d) from \cite{yue2021china}, which is above the  $3\times 10^{-9}\ m/s^2/Hz^{1/2}$ level from 1 mHz to 1 Hz and reach about $\sim 10^{-8}\ m/s^2/Hz^{1/2}$ in the mHz band.    
Since the z-axis is more noisy, the accurate estimation of the best performance level of the IS system will not be trivial, which is beyond the scope of this paper and will be left for future works.

\begin{figure}[H]
    \centering
    \includegraphics[scale=0.35]{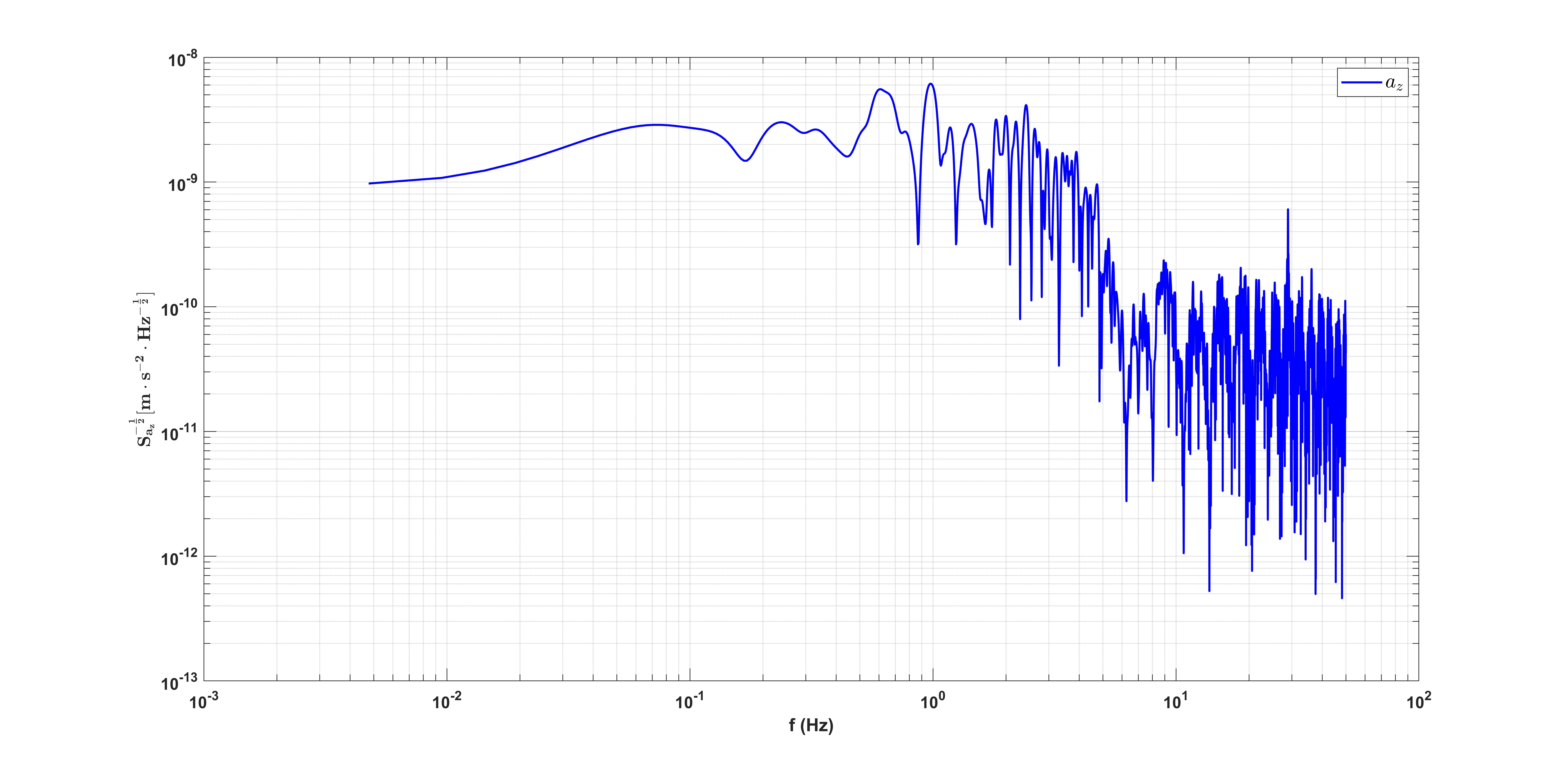}
    \caption{The noise floor of the actuation acceleration in z-axis after the systematic calibrations of the IS.}
    \label{fig:total_cal_result}
\end{figure}

Here, a data set of $V_{z1},\ V_{z2}$ in 2022 with good quality is selected for the performance evaluation. 
With the updated scale factors and biases, and after the COM corrections, the ASD curves of the TM linear accelerations $a_c^{z}$, which can be viewed as an estimation of the noise floor $a_{para,TM}^z$ in the z-axis measurement, is found in Fig. \ref{fig:total_cal_result}. 
It is found that, compared with the previous results, the new noise floor or resolution level is improved and now below the $3\times 10^{-9}\ m/s^2/Hz^{1/2}$ level in the sensitive band, and even below the $10^{-9}\ m/s^2/Hz^{1/2}$ level in the mHz band. This conclusion requires more analysis and further investigations, and is beyond the scope of this work. 

\section{Conclusion \label{Conclu}}

In this work, for gravity recovery missions, we suggest a systematic approach to calibrate the most important operation parameters for space-borne electrostatic IS system with two sets of satellite maneuvers. 
The operating parameters considered include the scale factors and COM offset vector that can be calibrated through one swing maneuver, and the intrinsic bias accelerations that can be calibrated through rolling maneuvers.
The time spans required for these satellite maneuvers depend on the accuracies of the specific IS payloads and the requirements on the precision of the parameter calibrations.
The working principle of the electrostatic IS and the estimation principles for its operating parameters are discussed, and the corresponding observation equations derived.

This systematic approach was applied to the re-calibrations of Taiji-1's IS system to study the drift and variations of the operating parameters after its two years operation. 
The two satellite maneuvers were conducted in MAY 2022 and { August} 2022 respectively. 
For Taiji-1, the short-time swing maneuver about 200 s is sufficient for scale factors and COM offset calibrations, and a data segment of a few thousand seconds long from the rolling maneuver is enough for a preliminary calibration of the biases.
The real rolling maneuver lasted for about 19 days, and this is for the accumulations of data segments with better qualities and for the improvement of the accuracy of the bias estimations.
The scale factors, COM offset and the intrinsic bias accelerations of Taiji-1's IS are precisely calibrated with the in-orbit data. 
The linear scale factor $k^y$ of the y-axis can not be calibrated by our new method, and its updated value is left for blank in this work.
The complete set of re-calibrated parameters are suggested to the Taiji-1 science team, and are archived in the Taiji-1 data center of CAS in Beijing for future use in data processing.

One of the main objectives of Taiji-1 was to evaluate the performance of the IS payload and test the related technologies.
Therefore, with the IS operating parameters being updated, a re-estimation of the performance of the IS and the comparison with the former results should be carried out. 
As mentioned, since the new scale factor for the best sensitive y-axis is not available, we re-processed the data of the z-axis and find that, compared with former results, the performance or noise floor is improved with the updated parameters. 
While a complete evaluation of the best performance of Taiji-1's IS system is not trivial, and this is beyond the scope of this paper and needs more in-depth investigations and analysis in future works.
On the other hand, in the extended phase in 2022, the Taiji-1 satellite operated in the high-low satellite-to-satellite tracking mode and could provide us monthly data of global gravity field.  
The re-calibration of the IS could improve the accuracies of the measurements of non-gravitational forces that disturb the orbit motions of the satellite, and the Taiji-1's global gravity model could be further updated. 

At last but not least, this systematic approach could offer high reference value for the ACC or IS calibrations of gravity recovery missions like the Chinese GRACE-type mission and the future planned Next Generation Gravity Missions.  
Also, it could shed some light on the in-orbit calibrations of the ultra-precision IS for future GW space antennas, since the principle and technology inheritance between these two generations of the electrostatic IS payloads.

% \indent This paper presents a systematic calibration approach for high-precision electrostatically suspended Inertial Sensors, successfully completing the calibration of the Taiji-1 Inertial Sensor and obtaining some improved results in the evaluation of its accuracy indicators. This method has important reference significance for satellite missions equipped with classic suspended Inertial Sensors, and also has technological inheritance for applications such as gravitational wave detection.  

% \indent In future work, the Taiji Scientific Collaboration will continue to improve this calibration system by combining on-orbit environmental optimization of maneuvering method with algorithmic iterations of data processing, achieving full-DOF high-precision calibration of onboard Inertial Sensors. It is expected that this system can be installed as a standardized module in relevant satellite projects to maintain the on-orbit precision of Inertial Sensors.

%%%%%%%%%%%%%%%%%%%%%%%%%%%%%%%%%%%%%%%%%%
%\section{Patents}

%This section is not mandatory, but may be added if there are patents resulting from %the work reported in this manuscript.

%%%%%%%%%%%%%%%%%%%%%%%%%%%%%%%%%%%%%%%%%%
\authorcontributions{P. Xu and H. Y.  Zhang designed the calibration experiments; D. Ye, L. E. Qiang, K. Q. Qi, S. X. Wang, Z. L. Wang and Z. R. Luo have discussed the principles and approved the feasibility of this calibration experiments for Taiji-1; Z. M. Cai uploaded the instructions to Taiji-1 and conducted the experiments; H. Y. Zhang, Z. Q. Ye analyzed the data and completed the calibrations; H. Y. Zhang and P. Xu composed the paper; P. Xu, L. E. Qiang, Z. R. Luo, J. G. Lei and Y. L. Wu provided general guidance and revised this paper.}

\acknowledgments{This work is supported by the National Key Research and Development Program of China No. 2020YFC2200601, No.  2020YFC2200602 and No. 2021YFC2201901, the Strategic Priority Research Program of the Chinese Academy of Sciences Grant No. XDA15020700, XDA15018000, and program No. ZC-1050-2021-05-011 from the Experiments for Space Exploration Program and the Qian Xuesen Laboratory, China Acadamy of Space Technology. The authors also acknowledge for the data resources from ``National Space Science Data Center, National Science \& Technology Infrastructure of China.''}

\conflictsofinterest{The authors declare no conflict of interest.}

%%%%%%%%%%%%%%%%%%%%%%%%%%%%%%%%%%%%%%%%%%
\begin{adjustwidth}{-\extralength}{0cm}

\reftitle{References}
\bibliography{ref1.bib}
\end{adjustwidth}
\end{document}